\documentclass[twocolumn,prd,preprintnumbers,nofootinbib,aps,floats,floatfix,amsmath,amssymb,secnumarabic]{revtex4}
\pdfoutput=1
\usepackage[usenames,dvipsnames]{color}
\usepackage[colorlinks, citecolor=blue]{hyperref}
\usepackage{bm,graphicx}
\usepackage{psfrag}
\usepackage[normalem]{ulem}

\def\be{\begin{equation}}
\def\ee{\end{equation}}    
\def\ba{\begin{eqnarray}}
\def\ea{\end{eqnarray}}

\def\lsim{\mbox{\raisebox{-.6ex}{~$\stackrel{<}{\sim}$~}}}
\def\gsim{\mbox{\raisebox{-.6ex}{~$\stackrel{>}{\sim}$~}}}

\def\zc{z_{\rm c}}
\def\delc{\delta_{\rm c}}
\def\delcx{\delta_{{\rm c},X}}
\def\nuc{\nu_{\rm c}}
\def\nucx{\nu_{{\rm c},X}}
\def\M{\mathcal{M}}
\def\Mh{\mathcal{M}^{\rm h}}
\def\Mf{\mathcal{M}^{\rm f}}
\def\ns{n_\mathrm{s}}

\newcommand{\del}[1]{\delta_{\rm c,#1}}

\begin{document}

\title{Number Counts and Non-Gaussianity}
\author{Sarah Shandera,$^1$}
\email{shandera@gravity.psu.edu}
\author{Adrienne L. Erickcek,$^{2,3}$}
\email{erickcek@cita.utoronto.ca}
\author{Pat Scott$^4$}
\email{patscott@physics.mcgill.ca}
\author{Jhon Yana Galarza$^{3,5}$}
\email{yanajho@gmail.com}
 \affiliation{  $^1$ Institute for Gravitation and the Cosmos, The Pennsylvania State University, University Park, PA 16802, USA\\
$^2$ Canadian Institute for Theoretical Astrophysics, University of Toronto, 60 St.~George Street, Toronto, Ontario M5S 3H8, Canada\\
$^3$ Perimeter Institute for Theoretical Physics, 31 Caroline St. N, Waterloo, Ontario N2L 2Y5, Canada\\
$^4$ Department of Physics, McGill University, Montr\'eal QC H2W2L8, Canada\\
$^5$ Universidad Nacional Mayor de San Marcos, Seminario Permanente de Astronom\'ia y Ciencias Espaciales, Facultad de Ciencias F\'isicas, Av. Venezuela s/n, Lima 1, Per\'u.}

\begin{abstract}
We describe a general procedure for using number counts of any object to constrain the probability distribution of the primordial fluctuations, allowing for generic weak non-Gaussianity. We apply this procedure to use limits on the abundance of primordial black holes and dark matter ultracompact minihalos (UCMHs) to characterize the allowed statistics of primordial fluctuations on very small scales. We present constraints on the power spectrum and the amplitude of the skewness for two different families of non-Gaussian distributions, distinguished by the relative importance of higher moments. Although primordial black holes probe the smallest scales, ultracompact minihalos provide significantly stronger constraints on the power spectrum and so are more likely to eventually provide small-scale constraints on non-Gaussianity.
\end{abstract}

\preprint{IGC-12/11-4}
\maketitle

\section{Introduction}
The power spectrum of primordial fluctuations is well measured on cosmological scales using the Cosmic Microwave Background (CMB) and Large Scale Structure \cite{Komatsu:2010fb,Dunkley:2010ge,Keisler:2011aw, Reid:2009xm, Sehgal:2010ca, McDonald:2004eu}. These observations provide compelling evidence that the fluctuations originated in an era of inflation. However, there is a great deal of new information waiting to be accessed in small-scale fluctuations and in higher-order statistics (non-Gaussianity). In this paper we explore both new regimes by examining how object number counts constrain the probability distribution of primordial fluctuations. By looking at very small objects (primordial black holes and ultracompact minihalos), we constrain the power in fluctuations on smaller scales than the CMB and Large Scale Structure currently probe, $k \gsim 3$ Mpc$^{-1}$. Since these objects are also very rare, we constrain the level of non-Gaussianity by limiting the abundance of extreme primordial overdensities.

Counts of rare objects are a useful probe of the primordial inhomogeneities in the gravitational field and their evolution: different objects probe different scales, different cosmological eras, and different particle physics and astrophysics. Primordial black holes (PBHs) have been used as a probe of small-scale power for many years \cite{Carr:1975qj}. Their abundance limits the allowed fluctuation power on very small scales, but the constraint is weak compared to the numbers expected from extrapolating a scale-invariant spectrum down from CMB scales \cite{Josan:2009qn,Carr:2009jm}. More recently, a much stronger constraint on small-scale power was obtained by considering ultracompact minihalos (UCMHs) of dark matter \cite{Josan:2010vn,Bringmann:2011ut,Li12}. The only drawback of this approach is that the strongest limits \cite{Bringmann:2011ut} require dark matter to annihilate into Standard Model (SM) particles, which can then be sought using standard indirect dark matter detection techniques \cite{Scott:2009tu}. Furthermore, UCMH constraints on the small-scale power obtainable by microlensing \cite{Li12} are still stronger than PBH constraints, and apply even for non-annihilating dark matter. 

The abundance of PBHs was also considered early on as a probe of non-Gaussianity \cite{Bullock:1996at}. In a non-Gaussian distribution, the number of rare objects is different from the Gaussian expectation, and rarer objects are typically sensitive to higher moments of the distribution. Number counts pick up {\it any} deviation from Gaussianity in a model-independent way and provide complementary constraints to those on the individual correlation functions, such as the shape of the 3-point function in momentum space (the bispectrum). However, there are important limitations on what can be learned from number counts on both the analytic and observational sides. First, in general one only knows the non-Gaussian probability density function (PDF) approximately, usually in terms of a few of the lowest-order moments. This limits how far out onto its tail the distribution is known, and so limits the utility of looking at very rare objects to constrain particular models. Furthermore, the best controlled approximations to weakly non-Gaussian PDFs are asymptotic expansions, so for a given level of non-Gaussianity one trusts the expansion only so far out onto the tail. The greater the level of non-Gaussianity, the more limited the range of utility of the asymptotic expansion. On the observational side, one must know the amplitude of fluctuations (the variance of the distribution) very precisely in order to find sharp constraints on the level of non-Gaussianity. The relationship between mass and the observable signature of the objects must be also be known accurately and precisely. Finally, extremely rare objects are not necessarily expected to be present in any finite sample (e.g.\, in any single survey or even our Universe), which makes drawing conclusions from their absence, or from very small number statistics, difficult. Nevertheless, constraints on the abundance of rare objects currently provide the only probe of the primordial fluctuations on very small scales, $k \gsim 10^4$ Mpc$^{-1}$. On larger scales, $50$ Mpc$^{-1}\lesssim k\lesssim 10^4$Mpc$^{-1}$, $\mu$-type spectral distortions of the CMB also provide constraints \cite{Hu:1994bz,Chluba:2011hw,Chluba:2012we}. We will return to the complementarity of these probes in the conclusions.

Here we treat the analytic difficulties mentioned above carefully and develop a general prescription to use number counts to constrain weakly non-Gaussian primordial fluctuations. We apply this technique to derive bounds on non-Gaussianities with UCMHs and PBHs. Previous authors have used PBHs to constrain particular cases of the non-Gaussian PDF or particular inflation scenarios. Bullock and Primack \cite{Bullock:1996at}, followed by Green and Liddle \cite{Green:1997sz} and Ivanov \cite{Ivanov:1997ia}, considered potentials with a localized feature that can both boost power dramatically on some scales and generate strong non-Gaussianity because of mode-coupling near the sharp feature. Pina Avelino \cite{PinaAvelino:2005rm} worked out constraints on a $\chi^2$ distribution for the fluctuations. Hidalgo \cite{Hidalgo:2007vk} considered a non-Gaussian PDF from only the first term in the Edgeworth expansion ($p_1$ from Eq.~(\ref{EdgeworthHier})). Klimai and Bugaev \cite{Klimai:2012sf} constrained two-field inflation models. Most recently Byrnes et al \cite{Byrnes:2012yx} considered constraints on weak or strong non-Gaussianity of the `local' type \cite{Salopek:1990jq}, where the primordial curvature is $\mathcal{R}(x)=\mathcal{R}_G(x)+\frac{3}{5}f_{NL}\mathcal{R}_G^2(x)$, where $\mathcal{R}_G(x)$ is a Gaussian random field.  Young \& Byrnes \cite{Young:2013oia} and Kohri et al. \cite{Kohri:2012yw} went on to also consider non-Gaussianity in the curvaton scenario.

Our results generalize all previous work on weakly non-Gaussian models, demonstrating how constraints can be obtained from any object abundance and for any weakly non-Gaussian distribution. We use an extended Press-Schechter \cite{Press:1973iz} approach where the abundance of an object is determined by the overdensity required for its formation ($\delta_\mathrm{c} \equiv \delta\rho_\mathrm{c}/\bar{\rho}$), the mass variance smoothed over the region that forms the object ($\sigma_R$), and similarly smoothed higher-order moments of the density fluctuations. These variables can be straightforwardly computed for {\it any} object: one need only compute the collapse threshold and smoothed statistics (by propagating the primordial fluctuations forward to the appropriate time of formation using transfer functions and a growth factor). This method is a straightforward extension of previous work on an analytic, weakly non-Gaussian mass function for galaxy clusters \cite{LoVerde:2007ri}. However, here we provide more precise criteria for where the analytic expansions are reliable in terms of the rareness of the object (the threshold $\nu_\mathrm{c}\equiv\delta_\mathrm{c}/\sigma_R$) and the level of non-Gaussianity. In addition, we consider a wider range of non-Gaussian scenarios by parameterizing deviations from Gaussianity with both the value of the dimensionless skewness ($\mathcal{M}_3$) and a rule for how the higher moments scale with respect to the skewness. Most previous work assumes one particular choice, but physical mechanisms generating the primordial fluctuations do not all follow that pattern, and the distinction is important for interpreting constraints. By demonstrating the computational limits of the Edgeworth expansion, we also clarify why strongly non-Gaussian models must be evaluated on a case-by-case basis.  Although we focus on small-scale constraints from PBHs and UCMHS in this work, the techniques we develop are equally applicable to limits on the abundances of high-redshift clusters, which provide strong constraints on non-Gaussianity on larger scales \cite{Shandera:2013mha}. 

Section \ref{sec:ExpansionTechnique} develops our formalism for describing weakly non-Gaussian distributions and determining the abundance of objects when the primordial fluctuations are described by a non-Gaussian distribution. Sections \ref{sec:PBH} and \ref{sec:UCMH} apply this technique to PBHs and UCMHs respectively, culminating with our final constraint plots in Figure \ref{fig:UCMH+PBH} (which a casual reader can skip to directly).  We conclude by comparing current and future constraints on small scale power and non-Gaussianity on a range of scales from a variety of existing and proposed techniques. 

\section{Calculating number counts using weakly non-Gaussian distributions}
\label{sec:ExpansionTechnique}

\subsection{Non-Gaussian Probability Distributions}
Assuming that objects form from regions of the density field above some threshold, we need a probability distribution $P$ of perturbation amplitudes $\delta_R$ and a threshold $\delc$ above which to count the objects. Here $\delcx$ gives the minimum density contrast required for a perturbation to later collapse into an object of type $X\in\{\rm PBHs$, UCMHs$\}$;  $\del{PBH}\sim1/3$ \cite{Carr:1975qj, Niemeyer:1999ak}, while $\del{UCMH}$ is a scale-dependent quantity \cite{Bringmann:2011ut}, although typically $\mathcal{O}(10^{-2}-10^{-3})$ \cite{Ricotti:2009bs,Bringmann:2011ut}.  Assuming a flat universe, the fractional density of matter in objects of type $X$ is
\be
\label{eq:betaX}
\beta(\delcx)=2A\int_{\delcx}^{\infty}P(\delta_R)\;d\delta_R,
\ee
where $P$ is a function of quantities smoothed on the scale $R$ associated with objects of mass $M$. For a generic PDF, the factor $A$ can be determined by ensuring that counting all objects recovers the total mass density. Here we will just use the Press-Schechter factor of 2 ($A=1$) as a good enough approximation since we are only concerned with a subset of objects\footnote{This factor would be more important if one had a particular model for how the primordial statistics varied as a function of scale and combined constraints on that model from objects on several scales.}. In addition, we have approximated the integral as extending to infinity, even though for some rare objects (e.g. UCMHs) that are not quite as rare as others (e.g. PBHs) there will be a finite upper limit associated with $\delc$ of the `next rarer' object.  Extending the integral to infinity is a very good approximation as long as $\delc$ is dissimilar for different rare objects, as $P$ is always a steeply-falling function of $\delta_R$.  If the probability distribution is Gaussian, we have
\be
\beta_{X,{\rm Gaussian}}(\nucx)={\rm erfc}\left[\frac{\nucx}{\sqrt{2}}\right],
\label{eq:betaXGauss}
\ee
where $\nucx\equiv\delcx/\sigma_R$ and $\sigma_R$ is calculated in the cosmological era relevant for formation of objects of type $X$.  In general we will suppress the subscript $X$ on $\beta$, $\delc$ and $\nuc$, except where talking about a specific type of object. Our convention follows \cite{Bringmann:2011ut}: $R$ refers to the comoving radius of the region that forms the object, $k = 1/R$ is the corresponding wavenumber, and both $\delcx$ and $\sigma_R$ are computed when the region enters the Hubble horizon. This is an analytic technique, which should be adjusted to agree with numerical simulations before it is trusted to give precise predictions of object abundance.

We can describe weakly non-Gaussian distributions as an expansion in moments about a Gaussian. A generic expression, generalizing the Edgeworth expansion, was written down by Petrov \cite{Petrov:1972, Petrov:1987}
\ba
\label{eq:Petrov}
P(\nu){\rm d}\nu&=&\frac{d\nu}{\sqrt{2\pi}}e^{-\nu^2/2}\left\{1+\sum_{s=1}^{\infty}\sigma_R^s\sum_{\{k_m\}}{H}_{s+2r}(\nu)\right.\nonumber\\
&&\left.\times\prod_{m=1}^s\frac{1}{k_m!}\left(\frac{S_{m+2,R}}{(m+2)!}\right)^{k_m}\right\}\;.
\ea
In this expression, the second summation is over all sets of integers $\{k_m\}$ that satisfy
\be
\label{eq:Diophantine}
k_1+2k_2+\dots+sk_s=s\;;
\ee
$H_n(\nu)$ are Hermite polynomials:\footnote{Note that there are two conventions for naming the Hermite polynomials, and these are sometimes denoted ${\rm He}_n$.} 
\be
H_n(\nu)=(-1)^ne^{\nu^2/2}\frac{{\rm d}^n}{{\rm d}\nu^n}e^{-\nu^2/2};
\ee
and $r=k_1+k_2+\dots+k_m$.  The ``reduced cumulants" are defined as
\be
S_{n,R}\equiv\frac{\langle\delta_R^n\rangle_c}{\langle\delta_R^2\rangle_c^{n-1}}\;.
\ee

To decide how to organize the terms in this series and truncate it somewhere, we need additional input about the relative importance of higher-order moments for the physical case under consideration. The choices are more naturally phrased in terms of the connected part of the dimensionless moments (the cumulants) of the real space density contrast, smoothed on scale $R$:
\be
\label{eq:MnRs}
\mathcal{M}_{n,R}=\frac{\langle\delta_R^n\rangle_c}{\langle\delta_R^2\rangle^{n/2}}\;.
\ee
The moments $\mathcal{M}_{n,R}$ can equally well be thought of as functions $\mathcal{M}_{n}(k)$ of the wavenumber $k$ corresponding to the smoothing scale $R$. We will use these two notations interchangeably, and sometimes we do not write out the scale-dependence explicitly at all.

Two choices motivated by particle physics \cite{Barnaby:2011pe} are hierarchical scaling
\be
\Mh_n \propto \left(\mathcal{I}\mathcal{P}^{\frac{1}{2}}_{\mathcal{R}}\right)^{n-2},
\ee
and feeder scaling
\be
\Mf_n \propto \;\mathcal{I}^{\,n}\;,\;\;\;n\geq3,
\ee
where $\mathcal{P}_{\mathcal{R}}$ is the amplitude of fluctuations in the primordial curvature, $\langle \mathcal{R}^2({\bf x}) \rangle = \int \frac{dk}{k} \mathcal{P}_\mathcal{R}(k)$, and $\mathcal{I}$ is a parameter indicating the strength of the interaction that sourced the non-Gaussianity.  In specific models one can work out the coefficients relating $\mathcal{I}\propto f_{NL}$, where $f_{NL}$ is the parameter for which CMB experiments quote constraints on non-Gaussianity. Using simple local models with each scaling to determine representative combinatorics\footnote{We use the local ansatz $\mathcal{R}(x)=\mathcal{R}_G(x)+ \frac{3}{5}f_{NL}[\mathcal{R}_G(x)^2-\langle\mathcal{R}_G(x)^2\rangle]$ for the hierarchical scaling (in that case $\mathcal{I}=f_{NL}$). We use a two field extension $\mathcal{R}(x)=\phi_G+\sigma_G+\frac{3}{5}\tilde{f}_{NL}[\sigma_G(x)^2-\langle\sigma_G(x)^2\rangle]$, with $\tilde{f}_{NL}\mathcal{P}_{\sigma}^{1/2}\gg 1$ for the feeder scaling (then $\mathcal{I}=\tilde{f}_{NL}^{eff}=\tilde{f}_{NL}\mathcal{P}_{\sigma}/\mathcal{P}_{\mathcal{R}}^{1/2}$).}, we can express the typical size of all higher order moments ($n\geq 3$) in terms of the third moment $\M_3$ by
\begin{align}
\label{eq:scaling}
\mathrm{Hierarchical:}\hspace{5mm} &\Mh_n = n!\,2^{n-3}\left(\frac{\Mh_3}{6}\right)^{n-2}\\
\mathrm{Feeder:}\hspace{5mm} &\Mf_n = (n-1)!\, 2^{n-1}\left(\frac{\Mf_3}{8}\right)^{n/3}
\end{align}
These are of course just two examples of how the moments could scale. For currently developed inflation models, they are representative of the full range of well motivated possibilities (which we discuss in more detail in the next section). Some models (quasi-single field inflation \cite{Chen:2009zp}) are a hybrid between the two scalings, or have additional numerical coefficients (e.g.\, two-field local models). Any skewness in $P(\nu)$ must be positive (assuming that higher moments are sub-dominant) for non-Gaussianities to increase the production of rare objects like PBHs, as the positive tail of the distribution must be enhanced relative to the core. 

Once the scaling of the moments has been established, Eq.~(\ref{eq:Petrov}) can be written in terms of the dimensionless smoothed moments $\mathcal{M}_{n,R}$, and the series can be organized in a form appropriate for any scaling. Making use of Eq.\ (\ref{eq:Diophantine}), we find
\begin{align}
\label{eq:PetrovM}
P(\nu){\rm d}\nu&=\frac{d\nu}{\sqrt{2\pi}}e^{-\nu^2/2}\left\{1+\sum_{s=1}^{\infty}\sum_{\{k_m\}}H_{s+2r}(\nu)\right.\nonumber\\
&\left.\times\prod_{m=1}^s\frac{1}{k_m!}\left(\frac{\mathcal{M}_{m+2,R}}{(m+2)!}\right)^{k_m}\right\}\nonumber\\
&\equiv\frac{d\nu}{\sqrt{2\pi}}e^{-\nu^2/2}(1+p_1(\nu,R)+p_2(\nu,R)+\dots)\;
\end{align}
For the two scalings above, we have
\ba
\label{EdgeworthHier}
p_1^{(h)}(\nu,R)&=&\frac{\mathcal{M}^h_{3,R}}{3!}H_3(\nu)\\\nonumber
p_2^{(h)}(\nu,R)&=&\frac{\mathcal{M}^h_{4,R}}{4!}H_4(\nu)+\frac{1}{2}\left(\frac{\mathcal{M}^h_{3,R}}{3!}\right)^2H_6(\nu)\\\nonumber
p_3^{(h)}(\nu,R)&=&\frac{\mathcal{M}^h_{5,R}}{5!}H_5(\nu)+\frac{\mathcal{M}^h_{3,R}\mathcal{M}^h_{4,R}}{3!4!}H_7(\nu)\\\nonumber
&&+\frac{1}{3!}\left(\frac{\mathcal{M}^h_{3,R}}{3!}\right)^3H_9(\nu)
\ea
for the hierarchical scaling (this is the usual Edgeworth expansion) and 
\ba
\label{EdgeworthAxion}
p_1^{(f)}(\nu,R)&=&\frac{\mathcal{M}^f_{3,R}}{3!}H_3(\nu)\\\nonumber
p_2^{(f)}(\nu,R)&=&\frac{\mathcal{M}^f_{4,R}}{4!}H_4(\nu)\\\nonumber
p_3^{(f)}(\nu,R)&=&\frac{\mathcal{M}^f_{5,R}}{5!}H_5(\nu)\\\nonumber
p_4^{(f)}(\nu,R)&=&\left(\frac{\mathcal{M}^f_{6,R}}{6!}+\frac{1}{2}\left(\frac{\mathcal{M}^f_{3,R}}{3!}\right)^2\right)H_6(\nu)\\\nonumber
p_5^{(f)}(\nu,R)&=&\left(\frac{\mathcal{M}^f_{3,R}\mathcal{M}^f_{4,R}}{3!4!}+\frac{\mathcal{M}^f_{7,R}}{7!}\right)H_7(\nu)
\ea
for the feeder scaling.  With this organization, the series in Eq.\ (\ref{eq:PetrovM}) is an asymptotic expansion for PDFs with moments of either scaling. Organizing the expansion in this way also exposes the key difference between the hierarchical and feeder scalings: $p^{(h)}_i \propto \mathcal{M}_3^i$, while $p_i^{(f)}\propto \mathcal{M}_3^{i/3}$, which implies that higher-order terms in the expansion make a larger contribution to  feeder distributions than they do to hierarchical distributions.

Then arranging the sums according to each scaling and performing the integral in Eq.\ (\ref{eq:betaX}), we find for objects of type $X$,
\ba
\label{eq:finalbeta}
\beta^{(h)}(\nu_{\mathrm{c},X})&=&{\rm erfc}\left(\frac{\nu_{\mathrm{c},X}}{\sqrt{2}}\right)\\\nonumber
&&+\,2\frac{e^{-\nu_{\mathrm{c},X}^2/2}}{\sqrt{2\pi}}\sum_{s=1}^{\infty}\sum_{\{k_m\}_h}{ H}_{s+2r-1}(\nu_{\mathrm{c},X})\\\nonumber
&&\times\prod_{m=1}^s\frac{1}{k_m!}\left(\frac{\mathcal{M}_{m+2,R}}{(m+2)!}\right)^{k_m};\\\nonumber
\beta^{(f)}(\nu_{\mathrm{c},X})&=&{\rm erfc}\left(\frac{\nu_{\mathrm{c},X}}{\sqrt{2}}\right)\\\nonumber
&&+\,2\frac{e^{-\nu_{\mathrm{c},X}^2/2}}{\sqrt{2\pi}}\sum_{s=1}^{\infty}\sum_{\{k_m\}_f}{H}_{s+1}(\nu_{\mathrm{c},X})\\\nonumber
&&\times\prod_{m=1}^s\frac{1}{k_m!}\left(\frac{\mathcal{M}_{m+2,R}}{(m+2)!}\right)^{k_m},
\ea
where the sets $\{k_m\}_h$ are again non-negative integer solutions to $k_1+2k_2+\dots+sk_s=s$ and $r=k_1+k_2+\dots+k_m$, but the sets $\{k_m\}_f$ are solutions to \mbox{$3k_1+4k_2+\dots+(s+2)k_s=s+2$}. Now for either scaling, truncating the series at some finite $s$ in the sums above keeps all terms up to the same order in $\mathcal{M}_3$: $\mathcal{M}^s_3$ for hierarchical scalings and $\mathcal{M}^{s/3}_3$ for feeder scalings.

\begin{figure*}[tbp]
\begin{center}
$\begin{array}{cc}
\includegraphics[width=0.5\textwidth,angle=0]{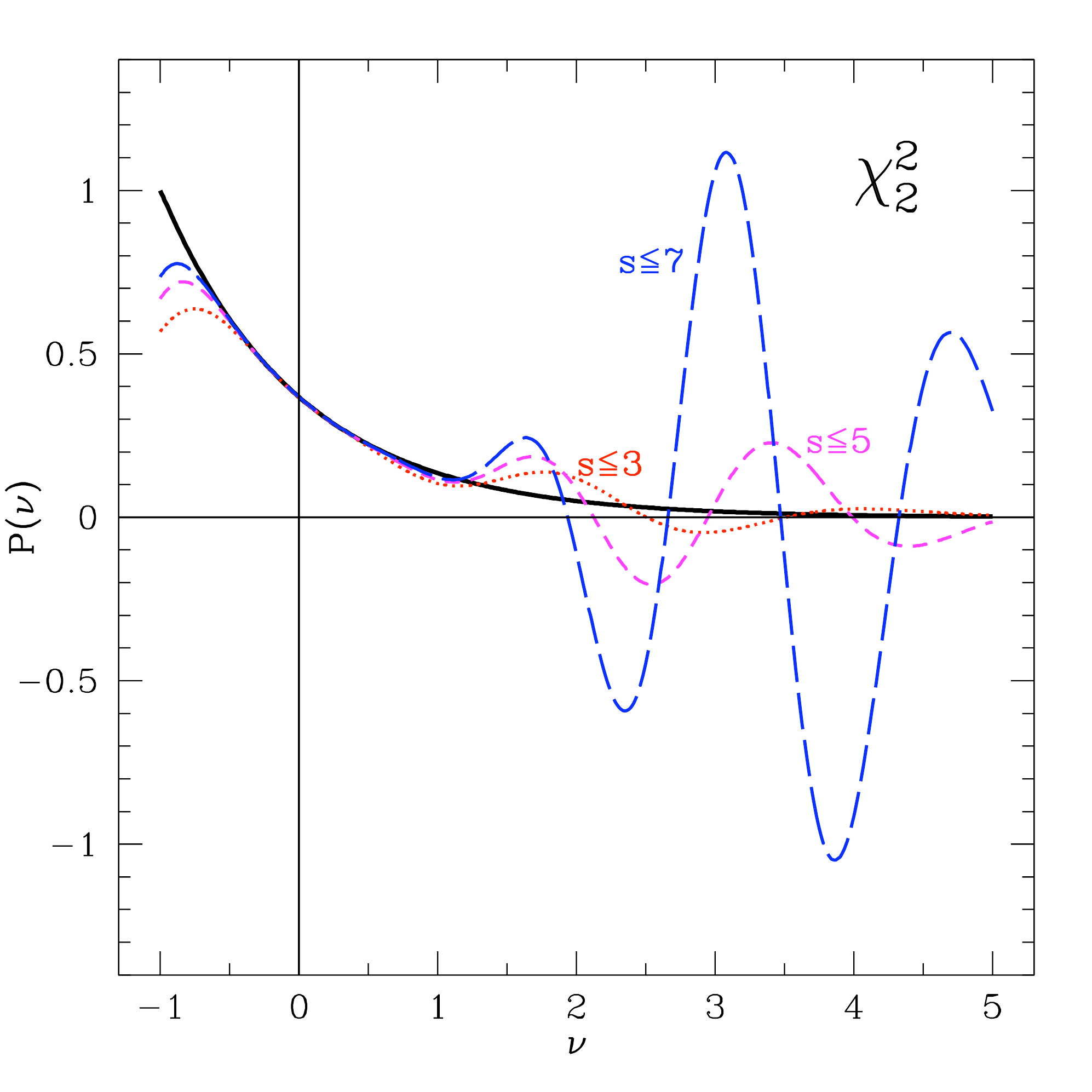} &
\includegraphics[width=0.5\textwidth,angle=0]{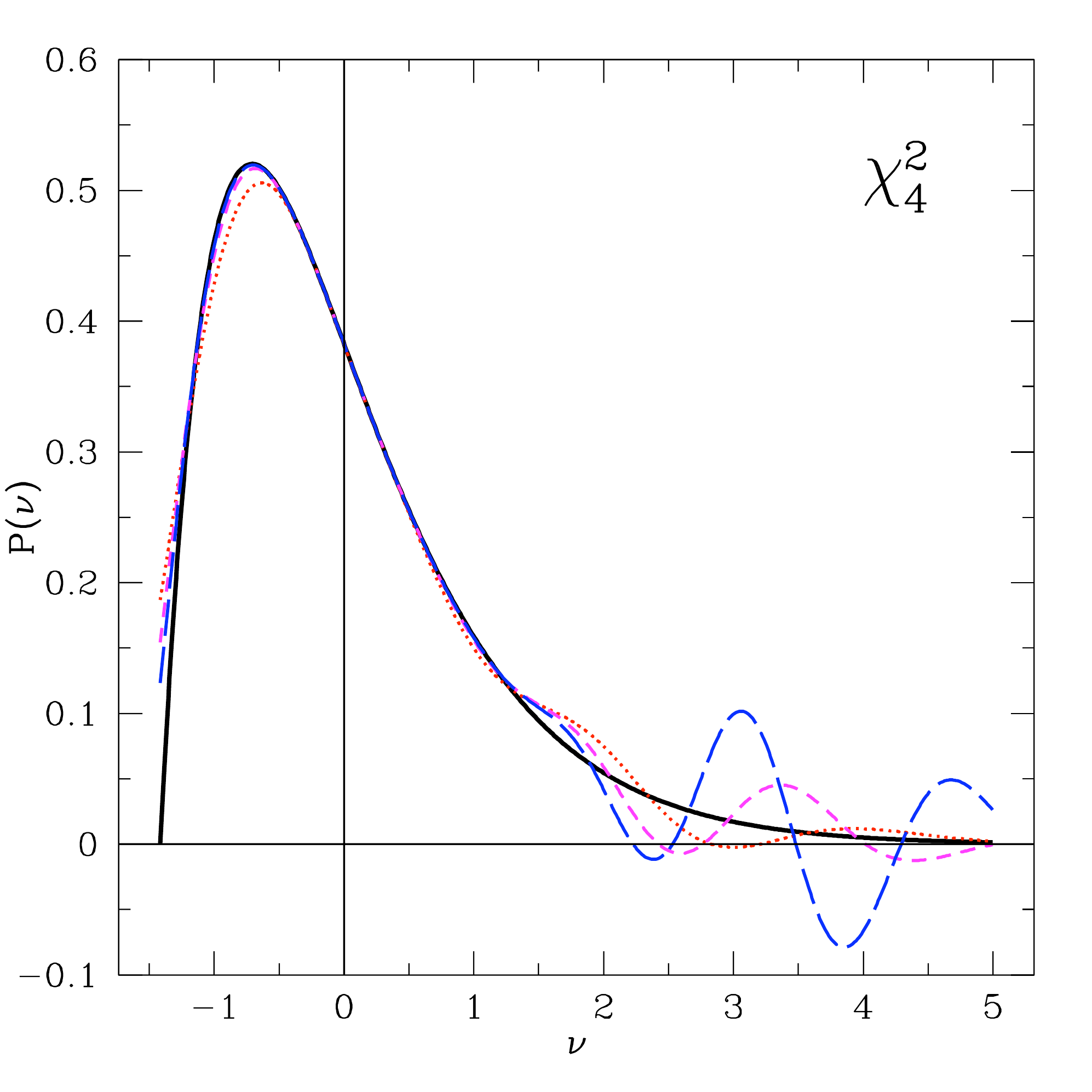} \\
\includegraphics[width=0.5\textwidth,angle=0]{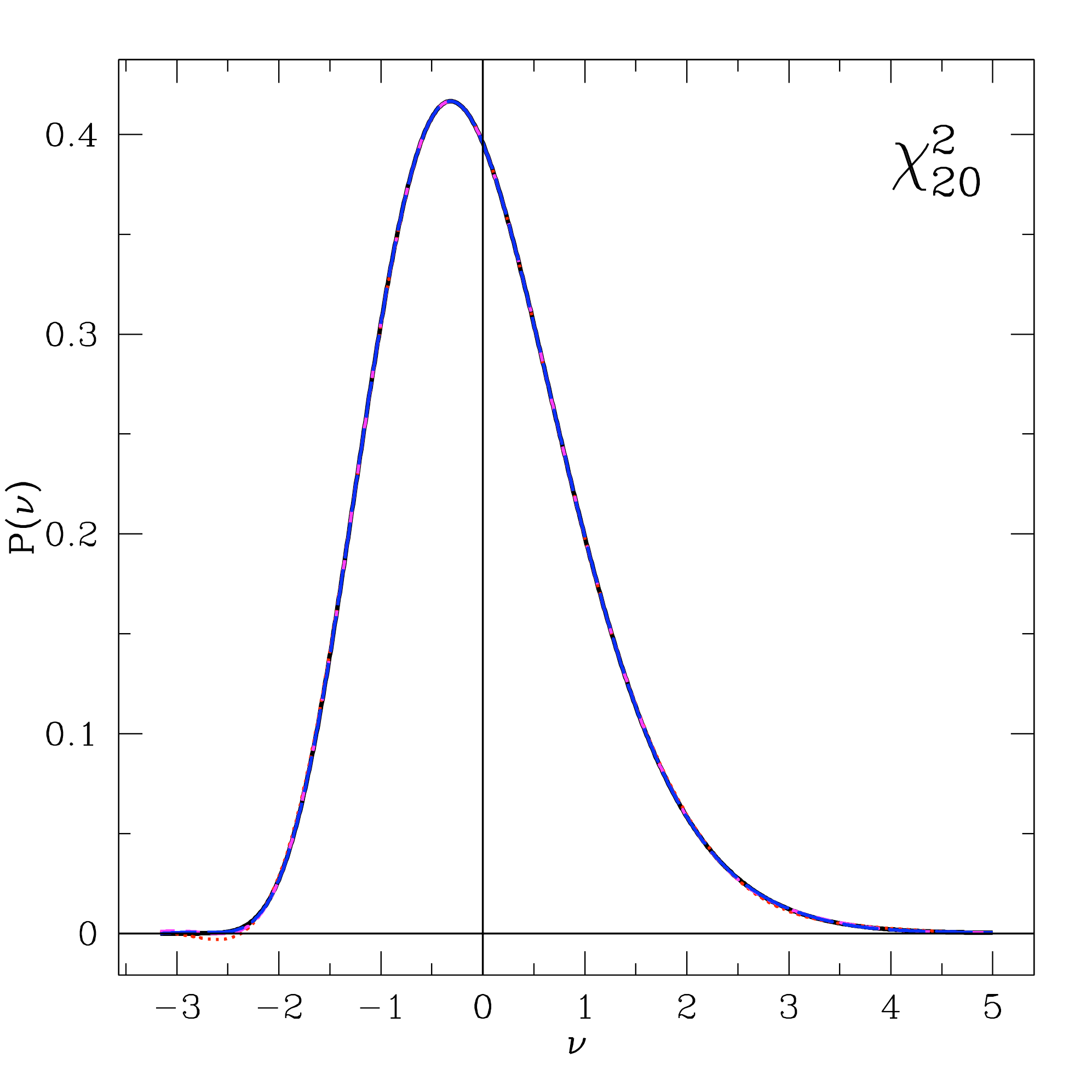} &
\includegraphics[width=0.5\textwidth,angle=0]{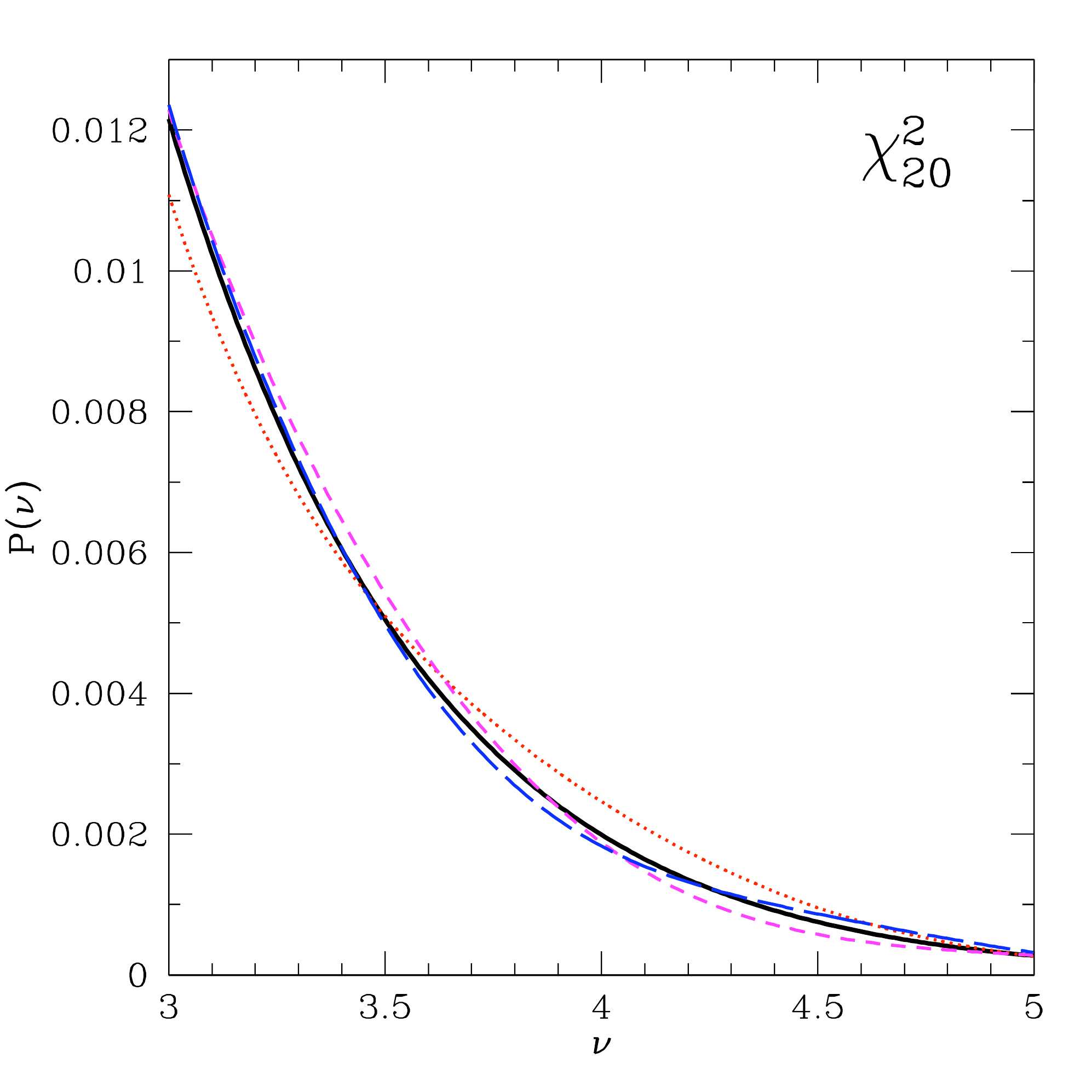} \\
\end{array}$
\caption{A comparison of the Edgeworth expansion to the exact PDF for the $\chi^2_k$ distribution with mean zero and variance one, $P(\nu)=\sqrt{2k}(\sqrt{2k}\,\nu+k)^{(k/2-1)}{\rm Exp}[-(\sqrt{2k}\,\nu+k)/2]/(2^{k/2}\Gamma(k/2))$. The upper left panel shows the full distribution (solid black line) for $k=2$, as well as the Edgeworth expansion truncated after $p_3$ (dotted red), $p_5$ (short-dashed pink), and $p_7$ (long-dashed blue), using hierarchical ordering. The upper right panel shows the same curves for $k=4$, while the lower panels are for $k=20$. The lower right is a zoomed-in view of the tail of the $k=20$ distribution. \label{fig:ChiSquare}}
\end{center}
\end{figure*}

We will work out constraints for the single parameter $\mathcal{M}_3$ characterizing the two classes of distribution (hierarchical and feeder) we have motivated using their particular Petrov expansions shown above. First, we illustrate how the Petrov-type expansions work in Figure \ref{fig:ChiSquare}, which shows a class of examples of the Edgeworth expansion compared to the corresponding complete distribution. The panels show three $\chi^2$ distributions with varying levels of non-Gaussianity and three truncations of the Edgeworth expansion for each case. As the figures show, the Edgeworth series is accurate out to larger values of $\nu$ when the distribution is closer to Gaussian. But, for a given level of non-Gaussianity, there is a value of $\nu$ beyond which the series is not an accurate fit and adding more terms does not improve the fit. A study comparing a variety of expansions of non-Gaussian distributions, with additional examples, can be found in \cite{Blinnikov:1997jq}.

\subsection{The approximate PDF and inflation models}
The parametrization of non-Gaussianity in terms of the dimensionless moments and their relative importance is the relevant organization for any observables sensitive to the PDF. There is a straightforward mapping between the dimensionless skewness and the power spectrum and bispectrum of any particular model (and choice of cosmological parameters). For example, for local and equilateral bispectral templates, the dimensionless skewness in the dark matter density perturbations at horizon crossing for modes that enter the horizon during radiation domination is
\begin{align}
\mathcal{M}_{3,R}&=3.13f_{\rm NL}^{\rm local}\mathcal{P}_{\mathcal{R}}^{1/2}(k\approx2R^{-1})\\
&=1.22f_{\rm NL}^{\rm equil}\mathcal{P}_{\mathcal{R}}^{1/2}(k\approx2R^{-1})
\end{align}
We have integrated over the bispectra, using a tophat window function and the transfer function described in the Appendix of \cite{Li12}. We have assumed that both the power spectrum and $f_{\rm NL}$ parameters are constant over the range where the integrals peak, $k\approx2/R$, which is not too restrictive since only about a decade in $k$ contributes significantly to the integrals. 

The dimensionless moments are straightforward to calculate for any inflation model, and the hierarchical and feeder scalings we use are representative of the full range of primordial non-Gaussianity that has its origin in the particle physics of inflation. Since the moments are determined from a perturbation-theory expansion of some Lagrangian, they do indeed scale in a simple way depending on a single parameter that determines the overall level of interaction together with order-one coefficients. Any scenario where the non-Gaussian interactions can be modeled by a single source, including inflaton self-interactions and curvaton models, gives rise to the hierarchical scaling. Scenarios where non-Gaussian fields couple to the primary source of curvature perturbation to generate the non-Gaussianity have the feeder scaling or a hybrid of the two types. Since a pure feeder scaling is the most non-Gaussian scaling (for a given skewness) known to be generated from inflation, the purely hierarchical and purely feeder scalings nicely bracket the behavior of all known inflation models. 

Finally, although we are dealing with much smaller scales than what is seen in the CMB, the behavior of the power spectrum and higher-order correlations on all scales can be computed in any given inflation model. So, the parameters constrained by probes of very small scales provide complementary constraints on the models. In addition, models where the non-Gaussianity changes on small scales typically have correlated changes in the amplitude of the power spectrum. The two effects are not arbitrary, as both come from the dynamics of the fields sourcing inflation. A full study of the relationship between the evolution of the power spectrum and amplitude of non-Gaussianity has not yet been done for a very wide variety of models, but see Figures 1 and 2 in \cite{Dias:2011xy} for an example\footnote{We thank D. Seery and R. Ribeiro for correspondence on this point.}.  Our joint constraints on the power spectrum and non-Gaussianity are particularly valuable for these models. We will return to this point when we present our results.

\subsection{Truncation and Error Evaluation}
\label{sec:Control}

To calculate the abundance of rare objects, we must truncate the $\beta$ expansions given by Eq.~(\ref{eq:finalbeta}), and we must determine the accuracy of the truncated series. For a given level of non-Gaussianity, there will be some region in the high-$\nu$ tail of the PDF for which the truncated expansion of the distribution is not reliable.  Furthermore, for a given value of $\nu$, there is also a level of non-Gaussianity beyond which the expansion is not reliable.  This maximum accessible value of $\mathcal{M}_3$ can be increased by including higher-order terms, but only up to a point.  Eventually the inclusion of more terms will stop improving the accuracy of the truncated sum, and adding additional terms makes the expansion less reliable (e.g., see the blue long-dashed curves in Figure \ref{fig:ChiSquare}).  In this section, we determine the optimal number of terms to include in the expressions for the PDF and $\beta$.   We also describe how we evaluate the accuracy of the resulting expressions and find the maximum value of $\mathcal{M}_3$ that can be reliably constrained by a given upper bound on $\beta$.  Some previous work on the validity of truncating this series for the hierarchical case only, and applying it to cluster number counts can be found in \cite{Chongchitnan:2010xz}. A recent study of observational constraints on the feeder and hierarchical scalings from X-ray detected clusters demonstrates that current data analysis is sensitive to the choice of scaling \cite{Shandera:2013mha}. 

To calculate the number of objects of a given type, we must be able to trust the PDF for values of $\nu>\nucx$ that contribute to the $\beta$ integral given by Eq.~(\ref{eq:betaX}).    Fortunately, only a limited range of $\nu$ values  make a significant contribution to $\beta$ because the PDF decreases rapidly as $\nu$ increases ($\sim \exp[-\nu^2/2]$).  Therefore, we only need to trust the PDF for $\nucx<\nu<\nu_\mathrm{max}$, where $\nu_\mathrm{max}$ is chosen so that truncating the $\beta$ integral at $\nu_\mathrm{max}$ has a minimal impact on $\beta$.  More explicitly, we demand that 
\be
\frac{\beta(\nu_\mathrm{max})}{\beta(\nucx)}< 0.02
\label{eq:vmaxcond}
\ee
so that the PDF for values of $\nu>\nu_\mathrm{max}$ contributes less than two percent of the $\beta$ integral.  If the observational upper bound on $\beta$ is greater than $5\times10^{-24}$, then Eq.~(\ref{eq:vmaxcond}) is satisfied for all relevant $\nucx$ values and all $\mathcal{M}_3$ values for which we can trust the PDF for $\nucx<\nu<\nu_\mathrm{max}$ if we set \mbox{$\nu_\mathrm{max} = K\nucx^{0.7}$}, where $K = 2.1$ for feeder models and $K = 2.2$ for hierarchical models.  Since all the bounds on UCMHs and most of the bounds on PBHs are less restrictive than $\beta < 5\times10^{-24}$, we adopt these expressions for $\nu_\mathrm{max}$ in our analysis.  Using stricter bounds on $\beta$ to constrain non-Gaussianity would require a larger value for $\nu_\mathrm{max}$, which could prevent these improved constraints on $\beta$ from providing improved constraints on $\mathcal{M}_3$ using asymptotic expansions of the PDF.

\begin{figure*}[tbp]
\begin{center}
$\begin{array}{cc}
\includegraphics[width=0.5\textwidth,angle=0]{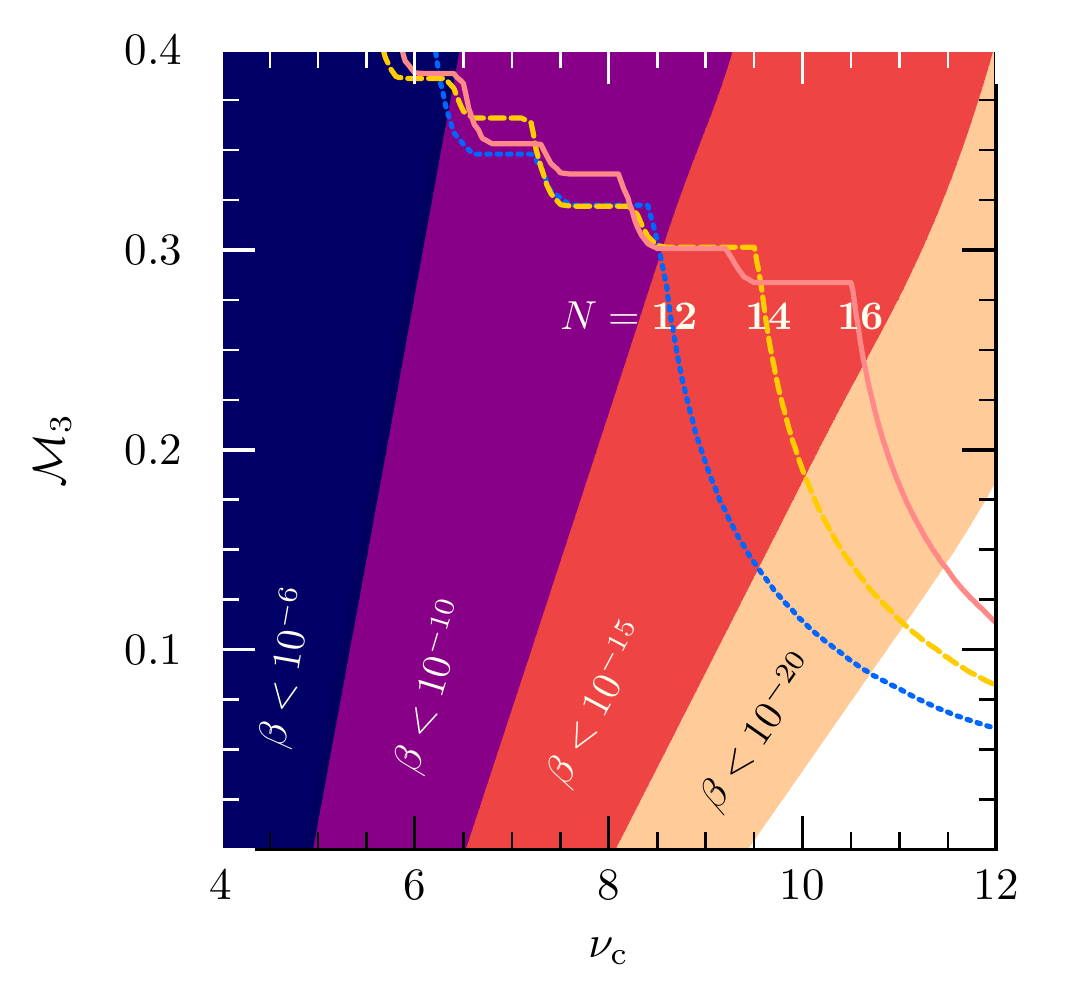} &
\includegraphics[width=0.5\textwidth,angle=0]{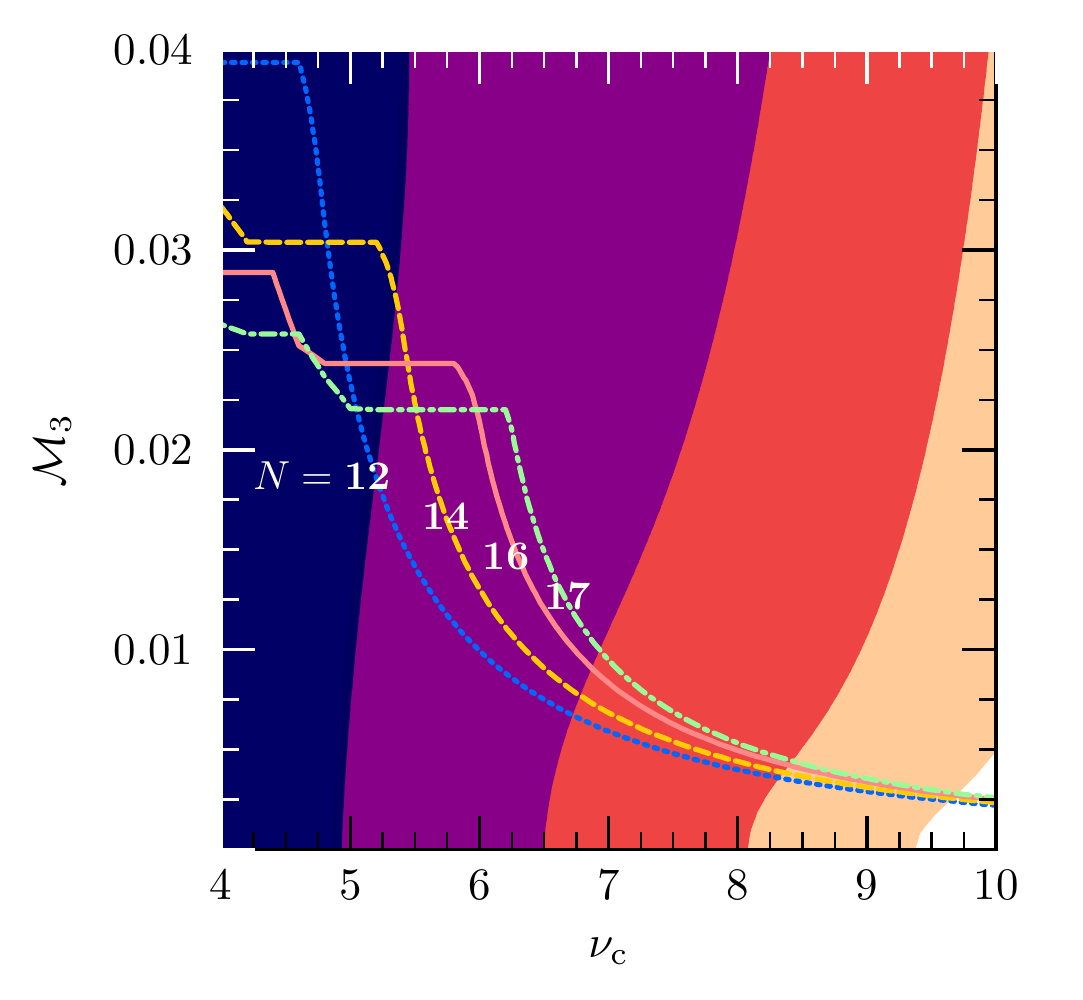} 
\end{array}$
\caption{Computational and observational constraints for hierarchical models (left) and feeder models (right).  The curves show the values of $\mathcal{M}_3$ above which the errors in the PDF truncated at $N$ terms exceed 20\% within the range of $\nu$ values that contribute to $\beta(\nucx)$.  The $N$ for each curve is indicated directly on the curves.  The shaded regions are excluded by observational constraints on $\beta$ with $N=16$ for hierarchical models and $N=17$ for feeder models, assuming $\beta< 10^{-6}$, $10^{-10}$, $10^{-15}$ or $10^{-20}$.  Only the parts of the shaded regions falling below the colored curves are reliably excluded.}
\label{procedure}
\end{center}
\end{figure*}

Next, we evaluate the error in the PDF over the range $\nucx<\nu<\nu_\mathrm{max}$ as a function of $\mathcal{M}_3$ and $\nucx$ .  Since the series expansion of the PDF is an asymptotic expansion, the error introduced by truncating the series at $N$ terms is the same order as the $N+1$ term in the expansion.\footnote{For both the hierarchical and the feeder models, we define the $N$th term in the expansion according to the power of $\mathcal{M}_3$ in that term.  For hierarchical models, the $N$th term is proportional to $\mathcal{M}_3^N$ and corresponds to the $s=N$ term in Eq.~(\ref{eq:PetrovM}).  For feeder models, the $N$th term is proportional to $\mathcal{M}_3^{N/3}$, and receives contributions from all terms with $s+2r=N$ in Eq.~(\ref{eq:PetrovM}).} 
Given $\mathcal{M}_3$ and $\nucx$, we compute the ratio of the $N+1$ term in the PDF to the sum of the first $N$ terms and find its maximum value in the range $\nucx<\nu<\nu_\mathrm{max}$: this ratio estimates the maximum fractional error in the truncated PDF over the range of $\nu$ values that contribute significantly to $\beta(\nucx)$.   

In our analysis, we restrict ourselves to values of $\mathcal{M}_3$ for which this maximum fractional error is less than twenty percent.   We use this criterion to define a ``computationally accessible" region in the $(\nucx,\mathcal{M}_3)$ parameter space.  Given that the PDF is accurate to within $20\%$ for all $\nu$ values that significantly contribute to $\beta(\nucx)$, we know that the fractional error in $\beta(\nucx)$ is at most $20\%$ for these values of $\mathcal{M}_3$ and $\nucx$, and we consider the truncated expansion to be trustworthy in this region.  We expect that the error in $\beta$ is actually lower than 20\% because the errors in the PDF are less than 20\% for most values of $\nu$ between $\nucx$ and $\nu_\mathrm{max}$.  Furthermore, these errors may be positive or negative and therefore may partially cancel when the PDF is integrated to obtain $\beta$. 

This expectation was confirmed when we applied our procedure to the $\chi_k^2$ probability distributions and compared the values of $\beta$ obtained by integrating the truncated PDF expansion over all $\nu>\nu_\mathrm{c}$ to the values obtained by integrating the true PDF (given in the caption to Figure \ref{fig:ChiSquare}).  We tested series that were truncated at $s=7,8,9,$ and $10$ and we considered the series trustworthy only if the estimated PDF errors were less than 20\% within the range $\nucx<\nu<\nu_\mathrm{max}$.  We estimated the errors in the PDF using the first term not included in the expansion; we did not compare the truncated PDF to the true PDF.  For all truncations, the $\beta$ values obtained from the truncated series with $\nu_\mathrm{c}$ and $k$ values in that truncation's trustworthy region differed from the true $\beta$ values by less than 6\%.
 
Figure \ref{procedure} shows the region of the $\nucx$-$\mathcal{M}_3$ plane that is computationally accessible for $N=12, 14,$ and $16$ for both hierarchical and feeder models, and $N=17$ for feeder models; the trustworthy regions lie below the thick curves. For hierarchical models, we see that adding terms with $N>12$ does not significantly change the upper bound on computationally accessible $\mathcal{M}_3$ values for $\nucx < 8$.  For larger values of $\nucx$, adding more terms to the PDF brings larger values of $\mathcal{M}_3$ into the trustworthy region.  For feeder models, adding more terms decreases the range of accessible $\mathcal{M}_3$ values for small values of $\nucx$, but improves the range for larger values of $\nucx$.  However, for $\nucx > 8$, adding more terms does not significantly change the maximum trustworthy value of $\mathcal{M}_3$.   

\begin{figure*}[tbp]
\begin{center}
$\begin{array}{cc}
\includegraphics[width=0.5\textwidth,angle=0]{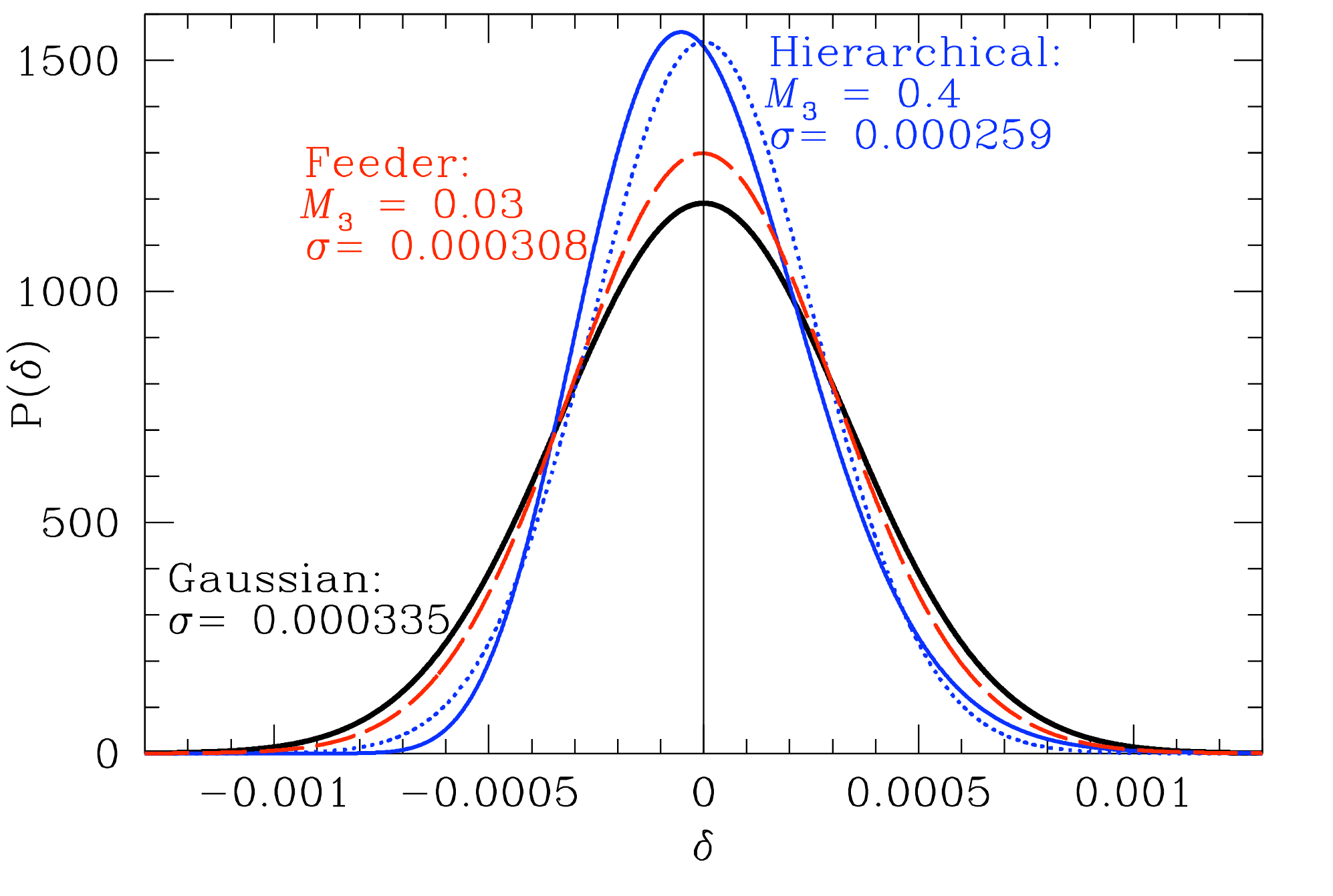} 
\includegraphics[width=0.5\textwidth,angle=0]{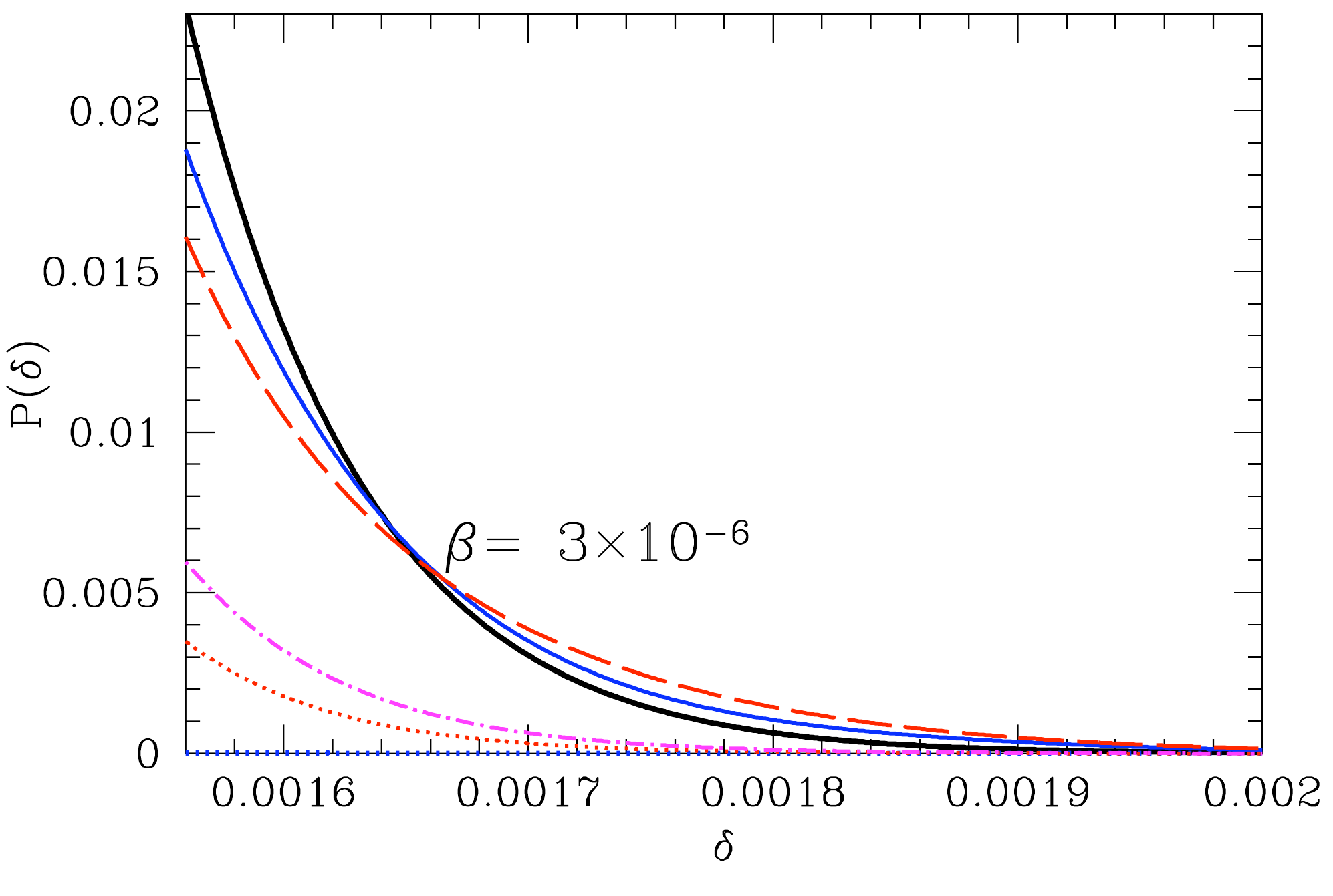} 
\end{array}$
\caption{Comparing Gaussian and non-Gaussian PDFs for UCMHs at $k=2.0\times10^7 \,{\rm Mpc}^{-1}$; the right panel shows $\delta \gtrsim \delta_\mathrm{c}\simeq1.56\times10^{-3}$.  In both panels, the thick solid black curve is the Gaussian PDF that saturates the observational bound on the UCMH abundance ($\beta \leq 3.0\times10^{-6}$).  Both panels also show the maximally non-Gaussian hierarchical PDF (solid blue) and feeder PDF (dashed red) that saturate this bound.  (That is, with the maximum value of $\mathcal{M}_3$ for which the expansion is controlled out to the largest value of $\nu$ that contributes significantly to the integral.) In the right panel, we indicate that all three of these curves have the same value of $\beta$.  The dotted curves show Gaussian PDFs that have the same values of $\sigma$ as these maximally non-Gaussian PDFs (blue for hierarchical and red for feeder).   In addition, the dot-dashed curve in the right hand panel shows a hierarchical PDF with the same values of $\sigma$ and $\mathcal{M}_{3}$ as the maximally non-Gaussian feeder PDF that saturates the $\beta$ bound (the red dashed curve).
\label{fig:PDFtails}}
\end{center}
\end{figure*}

To determine the optimal number of terms to consider when computing the PDF, we must know what $\nucx$ values we are interested in.  The shaded regions in Figure \ref{procedure} show the regions that are excluded by different observational upper bounds on $\beta$.   To maximize the power of a given upper bound on $\beta$, we want to choose $N$ to maximize the portion of the corresponding excluded region that lies within the computationally accessible region (i.e. beneath the thick curves in Figure \ref{procedure}).  For example, the darkest two shaded areas show the values excluded by the bound $\beta<10^{-10}$.  For hierarchical models, we should choose $N \geq 13$ when evaluating this constraint.  For feeder models $N=17$ is the best choice because it maximizes the excluded range of $\nucx$ values, but we could reach higher $\mathcal{M}_3$ by also considering PDFs with fewer terms.  In practice, however, changing the number of terms in the expansion for $\beta$ only changes the excluded regions for values of $\nucx$ near the excluded region's boundary and for untrustworthy $\mathcal{M}_3$ values that are far above all the thick curves in Figure \ref{procedure}. Therefore, we should choose $N$ such that the computationally accessible excluded region reaches the highest possible values of $\nucx$.  We can also trust this excluded region for all $\mathcal{M}_3$ that are computationally accessible for any smaller value of $N$.   Our default value for $N$ is 16 for hierarchical models and 17 for feeder models.  We chose this value of $N$ due to computational limitations, but from Figure \ref{procedure} we see that including more terms offers no benefits for hierarchical models with upper bounds on $\beta$ greater than $10^{-15}$.  For feeder models, including more terms could marginally extend the computationally accessible region excluded by $\beta<10^{-10}$, but it would not have any impact for lower or higher bounds on $\beta$.

Figure \ref{fig:PDFtails} compares several Gaussian and non-Gaussian PDFs (left panel) and their positive tails (right panel). The distributions have parameters found using the procedure above and the bound on the abundance of UCMHs at $k=2\times10^7 \,{\rm Mpc}^{-1}$, for which $\delta_\mathrm{c}\simeq0.00156$ and $\beta\lesssim3.0\times10^{-6}$. The thick solid black curve is the Gaussian distribution that saturates the bound on $\beta$, with $\sigma_R=0.000335$. The blue solid curve is the non-Gaussian distribution with hierarchical scaling and the maximum trustworthy value of $\mathcal{M}_3$ that gives $\beta=3.0\times10^{-6}$ ($\mathcal{M}_{3,{\rm max}}=0.4$ for $N=12$). This case also defines a minimum variance $\sigma_R=0.000259$ for a hierarchical distribution that saturates the bound on $\beta$; the dotted blue curve is the Gaussian distribution with that same value of $\sigma$.  In the right hand panel, this Gaussian distribution has very little area under the positive tail and lies nearly on top of the $x$-axis.  The long-dashed red curve is the non-Gaussian distribution with feeder scaling and the maximum trustworthy value of $\mathcal{M}_3$ that gives $\beta=3.0\times10^{-6}$ ($\mathcal{M}_{3,{\rm max}}=0.03$ for $N=13$).  This curve has $\sigma_R=0.000308$, and the dotted red curve in the right panel shows the tail of a Gaussian distribution with the same value of $\sigma$.  In the right panel, this curve would be nearly indistinguishable from the red dashed curve.  The dot-dashed curve in the right panel is one way of visualizing the difference between the feeder and hierarchical scalings: it shows a hierarchical non-Gaussian PDF with values of $\sigma$ and $\mathcal{M}_3$ that saturate $\beta$ when used in a {\it feeder} PDF ($\mathcal{M}_{3,{\rm max}}=0.03$ and $\sigma_R=0.000308$, as shown by the red dashed curve).  Clearly, for a particular value of $\mathcal{M}_3$, the feeder scaling gives a PDF that is overall substantially more non-Gaussian than the hierarchical scaling. 

\begin{figure*}[tb]
\begin{center}
$\begin{array}{cc}
\includegraphics[width=0.5\textwidth,angle=0]{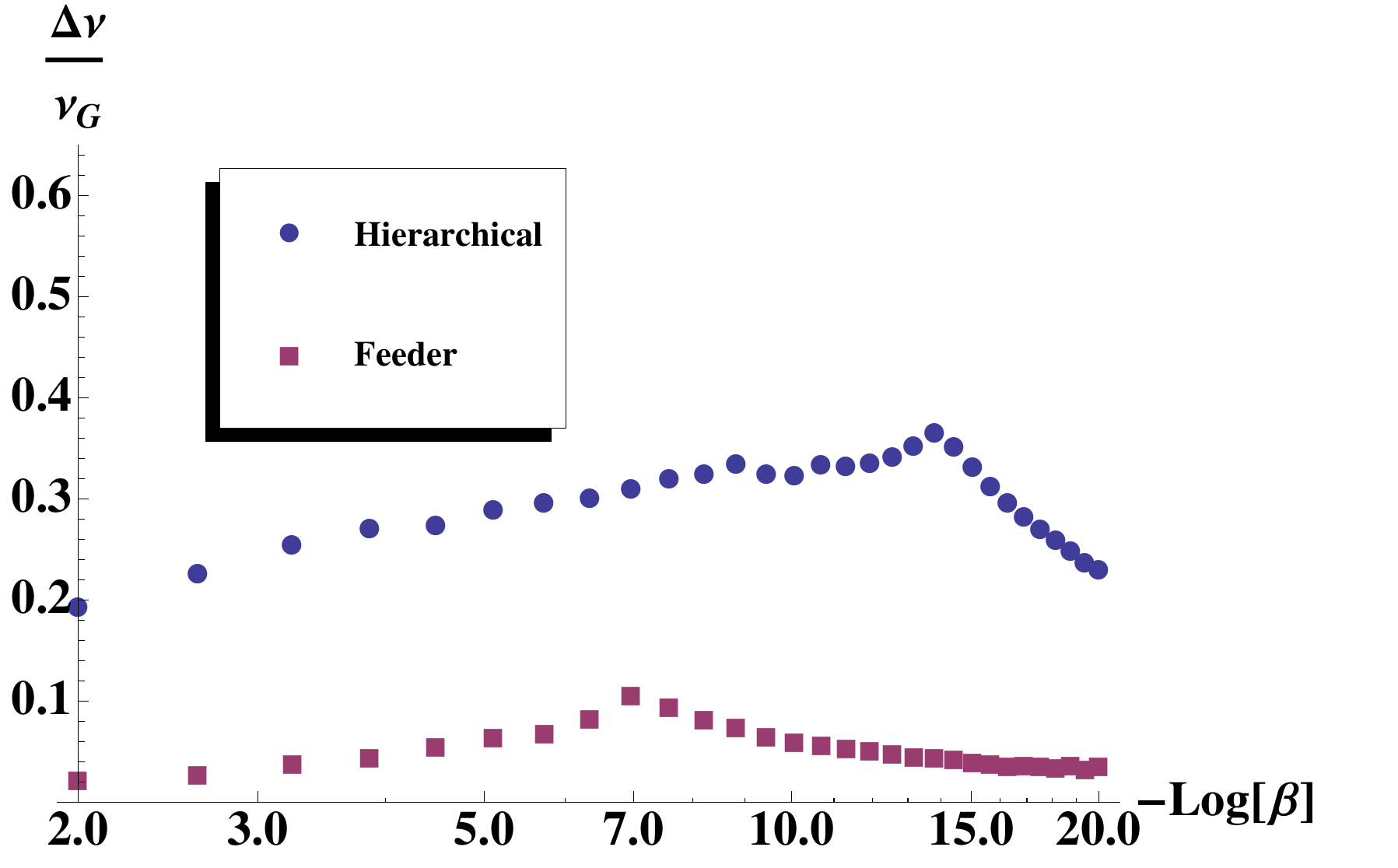} 
\includegraphics[width=0.5\textwidth,angle=0]{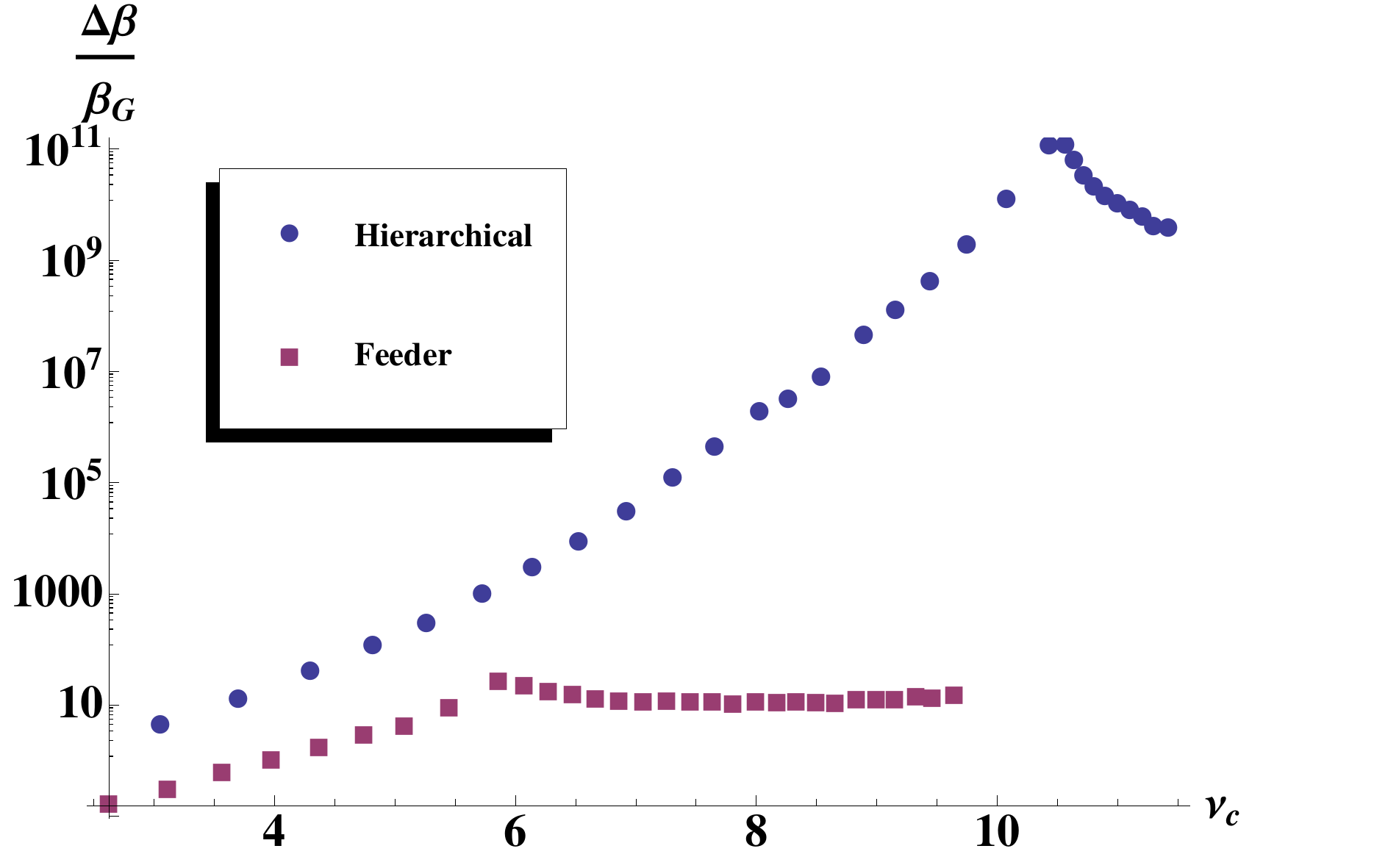} 
\end{array}$
\caption{{\bf Left panel:} The ratio $[\nu_\mathrm{c,NG}(\mathcal{M}_{3,{\rm max}})-\nu_\mathrm{c,G}]/\nu_\mathrm{c,G}$, where $\nu_\mathrm{c,G}$ is the value of $\nu_\mathrm{c}$ such that a Gaussian distribution gives the plotted value of $\beta$ (see Eq.\ (\ref{eq:betaXGauss})) and $\nu_\mathrm{c,NG}(\mathcal{M}_{3,{\rm max}})$ is the value of $\nu_\mathrm{c}$ such that a non-Gaussian distribution with $\mathcal{M}_{3} = \mathcal{M}_{3,\mathrm{max}}$ gives the same value of $\beta$.  Here $\mathcal{M}_{3,\mathrm{max}}$ is the largest amplitude of non-Gaussianity where the Petrov expansion, calculated out to $N=16$ for hierarchical scaling and $N=17$ for feeder scaling, is trustworthy, according to the criteria discussed below Eq.\ (\ref{eq:vmaxcond}).  {\bf Right panel:} The ratio of $[\beta(\mathcal{M}_{3,{\rm max}},\nu_\mathrm{c})-\beta(0,\nu_\mathrm{c})]/\beta(0,\nu_\mathrm{c})$ plotted versus $\nu_\mathrm{c}$ assuming hierarchical (blue circles) or feeder (red squares) scaling. In both plots, the change in slope in the hierarchical points appears to be an artifact of our truncation to a finite number of terms - going beyond $N=16$ would probably allow one to continue go farther onto the tail. For the feeder scaling, however, it is less clear that adding more terms will help - the distribution is quite non-Gaussian, so the Petrov expansion breaks down irreparably at smaller $\nu$ and smaller $\mathcal{M}_3$. The scatter in the points about a smooth line comes from the approximate nature of our criteria for determining $\mathcal{M}_{3,{\rm max}}$.}
\label{fig:deltanu}
\end{center}
\end{figure*}

Now we can get some measure of the maximum shift in the PDF that can come from adding weak non-Gaussianity in a controlled way. Figure \ref{fig:deltanu} illustrates this in two ways. The left hand panel illustrates how a constraint on the variance $\sigma$ derived assuming a Gaussian distribution can be shifted to a constraint on a non-Gaussian distribution with a new (smaller) variance. The vertical axis shows the relative shift in $\nu_\mathrm{c}$ between the Gaussian distribution that gives abundance fraction $\beta$ and the maximally non-Gaussian distribution, under control according to our criteria above, that gives the same $\beta$. To use this for any particular object, $\nu_\mathrm{c}$ should be written in terms of the appropriate $\delta_\mathrm{c}$ and $\sigma_R$ for the object of interest. $\mathcal{M}_{3,{\rm max}}$ can be likewise be converted to a non-Gaussian parameter in the primordial distribution.  This plot gives a sense of how accurately the variance of fluctuations must be known in order for number counts to provide an independent constraint on non-Gaussianity using Petrov expansions.  For instance, if $\Delta\nu/\nu_G = 0.3$ in the left panel of Figure \ref{fig:deltanu}, then adding weak non-Gaussianity in a controlled way has the same effect on the abundance of rare objects as increasing $\sigma_R$ by $\sim\!\!30\%$.  If the uncertainty in $\sigma_R$ exceeds $30\%$, then number counts with the corresponding value $\beta$ cannot independently constrain non-Gaussianity without knowing the full PDF of the density fluctuations.

The right hand panel of  Figure \ref{fig:deltanu} shows the maximum fractional change in the abundance $\beta$ that results from adding controlled non-Gaussianity for a given object and fixed variance (so fixed $\nu_\mathrm{c}$): $[\beta(\mathcal{M}_{3,\mathrm{max}},\nu_\mathrm{c})-\beta(0,\nu_\mathrm{c})]/\beta(0,\nu_\mathrm{c})$. Notice that the primary limit on how much $\beta$ can change is our ability to accurately determine the PDF. Since the feeder scaling produces a distribution that is overall much more non-Gaussian than the hierarchical for a fixed value of $\mathcal{M}_3$, we cannot accurately calculate at as large of $\nu_\mathrm{c}$ (rarer objects) or for as large of shifts in $\beta$ as for the hierarchical case. Again, the effects of more non-Gaussian distributions can be computed if one knows the full distribution rather than just a finite number of moments.

\section{Primordial Black Holes}
\label{sec:PBH}
For a Gaussian spectrum of fluctuations, the present-day fractional density of black holes has typically been used to constrain the spectral index $\ns$ under the assumption of a scale-free spectrum \cite{Bullock:1996at, Green:1997sz, Green:2004wb, Chongchitnan:2006wx}.  The resulting bound, $\ns\lesssim 1.3$ (based on e.g. VERITAS searches at $k=3\times10^{15}$\,Mpc$^{-1}$ \cite{Tesic:2012kx}), is easily satisfied by the current experimentally measured value $\ns=0.9603\pm 0.0073$ \cite{PlanckInflation}, with little or no running.  In terms of generalized, scale-dependent Gaussian power spectra, PBHs have been used to place bounds on the amplitude of curvature perturbations at the level of $\mathcal{P}_\mathcal{R}\lesssim10^{-2}$ \cite{Josan:2009qn,Kaz12}, on scales corresponding to $10^{-2}<k\le10^{19}$\,Mpc$^{-1}$.  Stronger constraints than this can be found if the analysis is done using a top-hat window function rather than a Gaussian, which is arguably a more correct method \cite{Bringmann:2001yp,Bringmann:2011ut}.

Present observational bounds on PBHs \cite{Josan:2009qn,Carr:2009jm} limit $\beta_\mathrm{PBH}$ to values ranging from $\beta_\mathrm{PBH}\lesssim 10^{-10}$ for solar mass PBHs, to $\beta_\mathrm{PBH}\lesssim 10^{-28}$ for a small range of PBH masses around $10^{13}$\,g, to $\beta_\mathrm{PBH}\lesssim 10^{-20}$ for even smaller masses.  A PBH mass $M_\mathrm{PBH}$ corresponds to horizon mass $M_\mathrm{H}=3^{3/2} M_\mathrm{PBH}$ \cite{Josan:2009qn}, and therefore wavenumber
\begin{equation}
\frac{k}{k_\mathrm{eq}} = (2\times3^3)^{-1/4} \left(\frac{g_{\star}^R}{g_{\star}^\mathrm{eq}}\right)^{1/4} \left(\frac{g_{\star S}^\mathrm{eq}}{g_{\star S}^R}\right)^{1/3} \left(\frac{M_\mathrm{H}^\mathrm{eq}}{M_\mathrm{PBH}}\right)^{1/2}.
\end{equation}
\noindent Here $k_\mathrm{eq}=0.072\Omega_\mathrm{m}h^2$\,Mpc$^{-1}=9.68\times10^{-3}$\,Mpc$^{-1}$ \cite{Komatsu:2010fb} and $M_\mathrm{H}^\mathrm{eq}=3.5\times10^{17}\,M_\odot$ \cite{Scott:2009tu} are the wavenumber and horizon mass respectively at equality, and $g_{\star S}^\mathrm{eq}=3.91$, $g_{\star}^\mathrm{eq}=3.36$, $g_{\star}^R\ge106.75$ \cite{KolbTurner} are the respective degrees of freedom at equality and the time of horizon entry.  

We use the current limits on $\beta_\mathrm{PBH}$ from \cite{Carr:2009jm}, along with Eq.~(\ref{eq:finalbeta}) and the requirement that it represents a well-behaved expansion, to determine the range of values of $\M_3(k)$ that are excluded by non-observation of PBHs.  For this calculation we draw upon the expressions for $\sigma_R$ in a generalized Gaussian power spectrum, given in Appendix B of \cite{Bringmann:2011ut}; in particular, we employ the top-hat window function advocated in that paper.  We give our limits in Figure \ref{fig:UCMH+PBH} as a function of the power spectrum of curvature perturbations at any given $k$:
\begin{equation}
\label{Pgauss}
\mathcal{P}_\mathcal{R}(k) = 1.10\,\sigma_R^2.
\end{equation}
We show the excluded areas as dark shaded regions, for both hierarchical and feeder scaling, and at three different example scales.  The corresponding limits on $\beta$ at these scales are $\beta < 2.2\times10^{-13}$ ($k=2\times10^{3}$\,Mpc$^{-1}$), $\beta < 9.4\times10^{-12}$ ($k=2\times10^{7}$\,Mpc$^{-1}$) and $\beta < 1.4\times10^{-23}$ ($k=2\times10^{17}$\,Mpc$^{-1}$).  More positive non-Gaussianity (larger $\M_3$) results in stronger exclusions, up until the point at which our expansion Eq.~(\ref{eq:finalbeta}) begins to break down.  Above these very large values of $\M_3$, the excluded regions shrink with additional non-Gaussianity, as more of the available parameter space is made inaccessible by the increasingly poor behavior of the expansion.

\section{Ultracompact Minihalos}
\label{sec:UCMH}
\begin{figure*}[p]
\begin{center}
\includegraphics[width=0.4\textwidth]{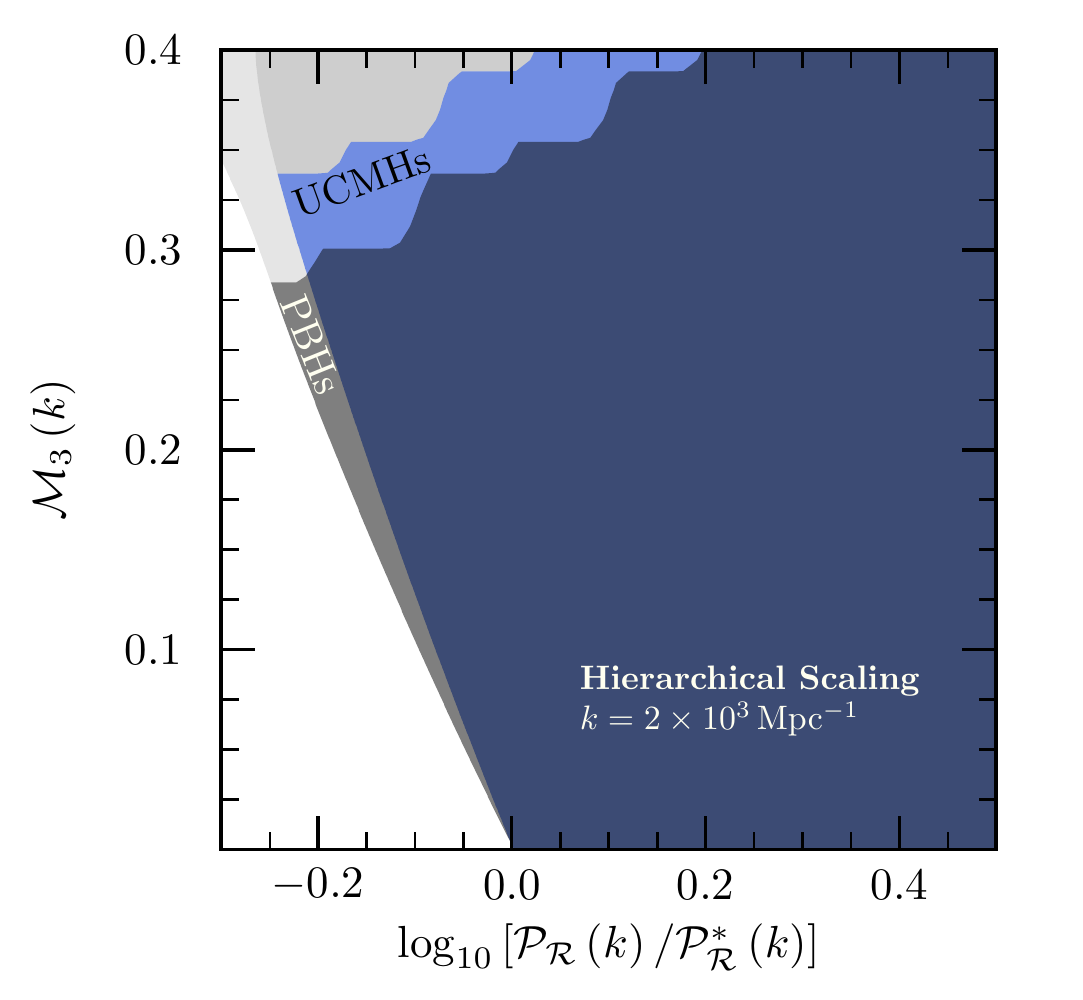}%
\includegraphics[width=0.4\textwidth]{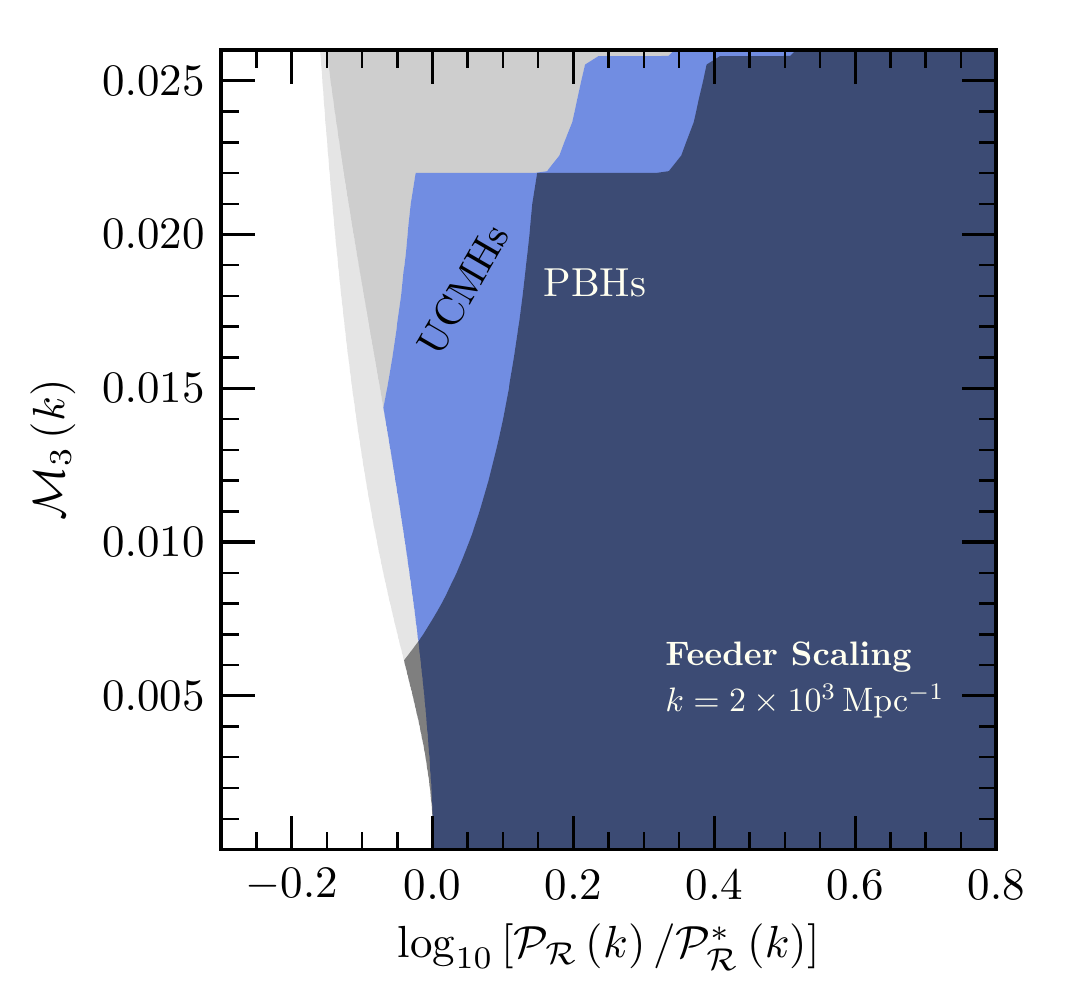} 
\includegraphics[width=0.4\textwidth]{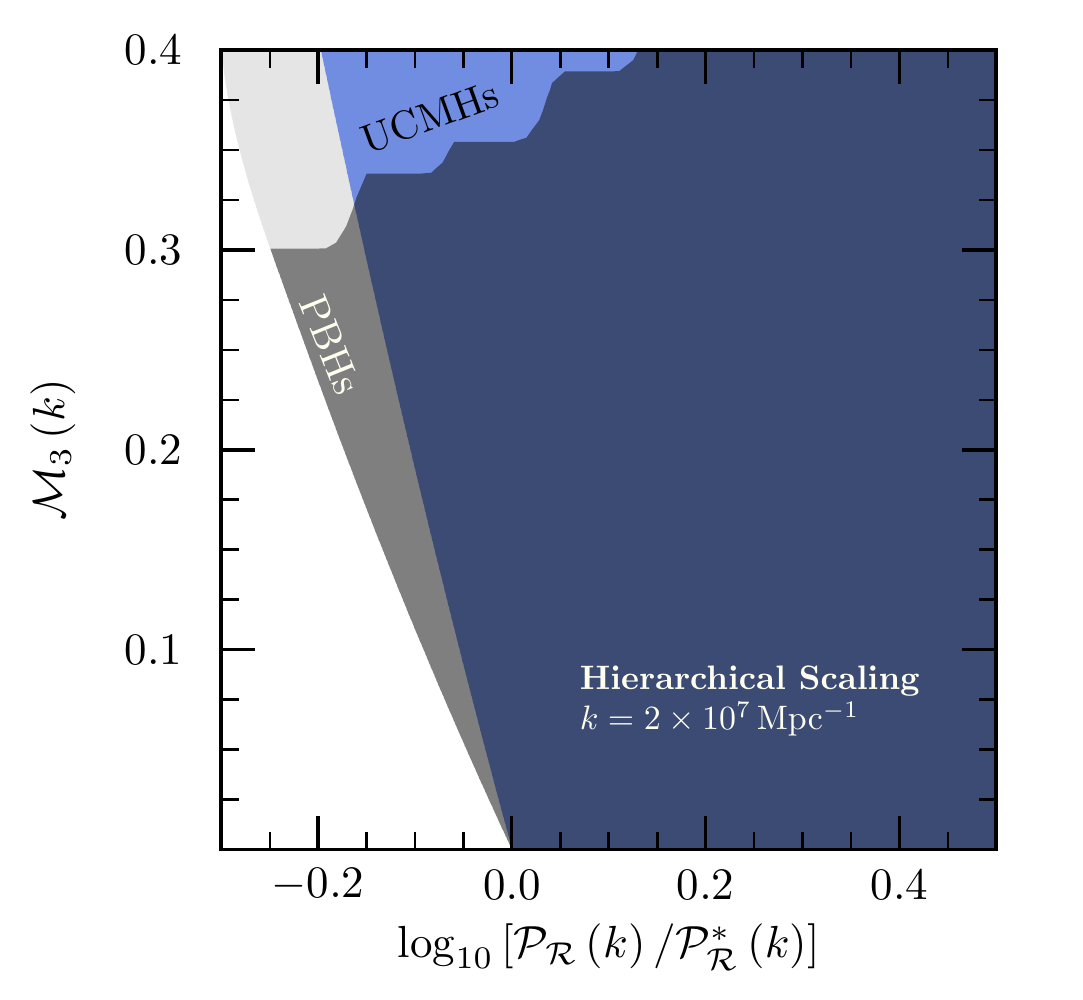}%
\includegraphics[width=0.4\textwidth]{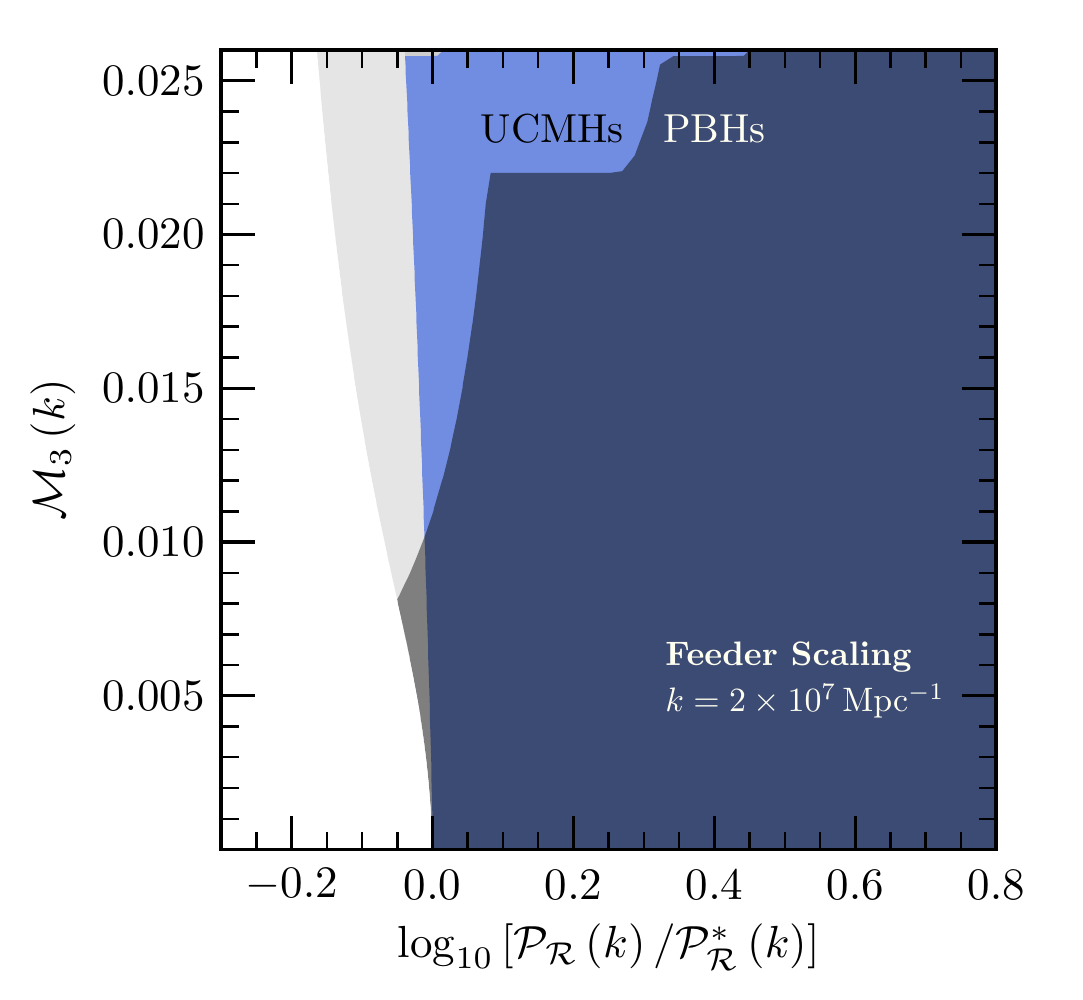} 
\includegraphics[width=0.4\textwidth]{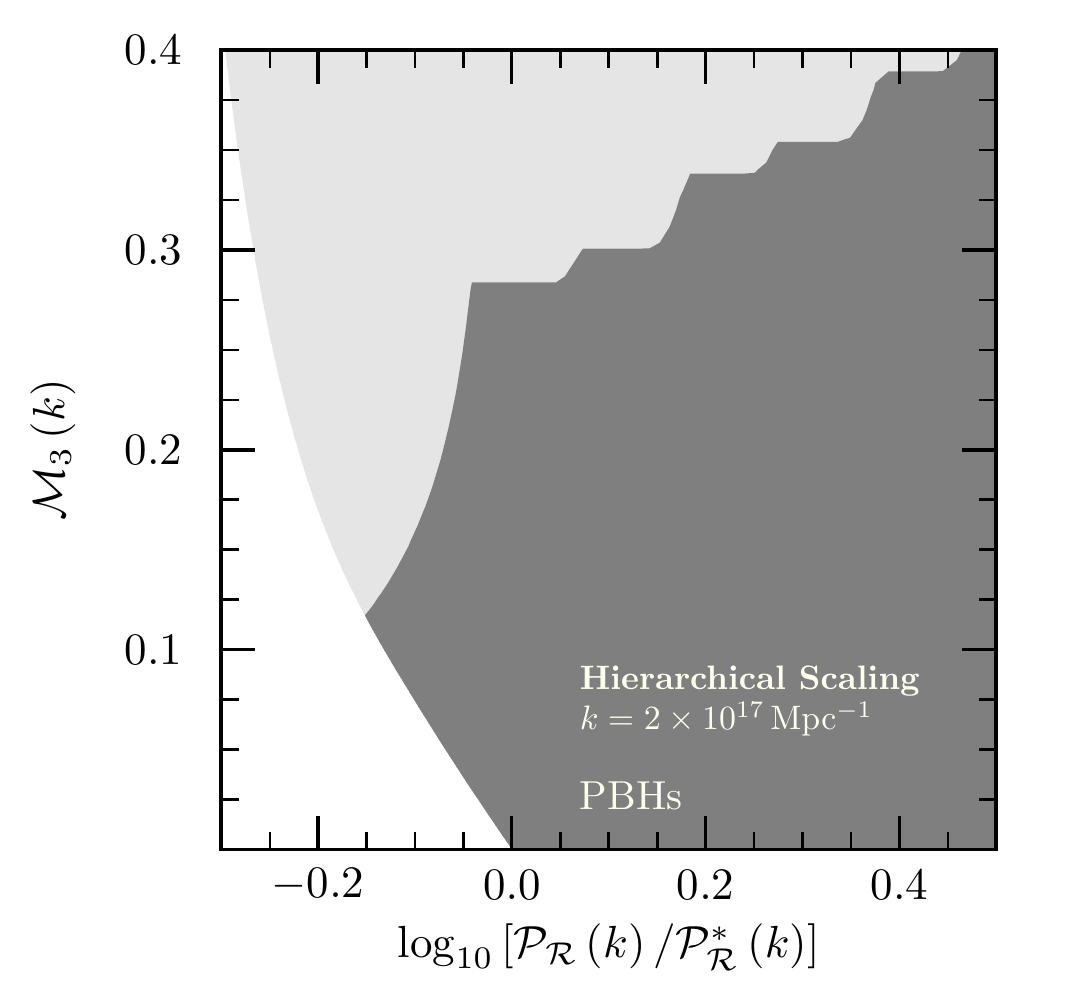}%
\includegraphics[width=0.4\textwidth]{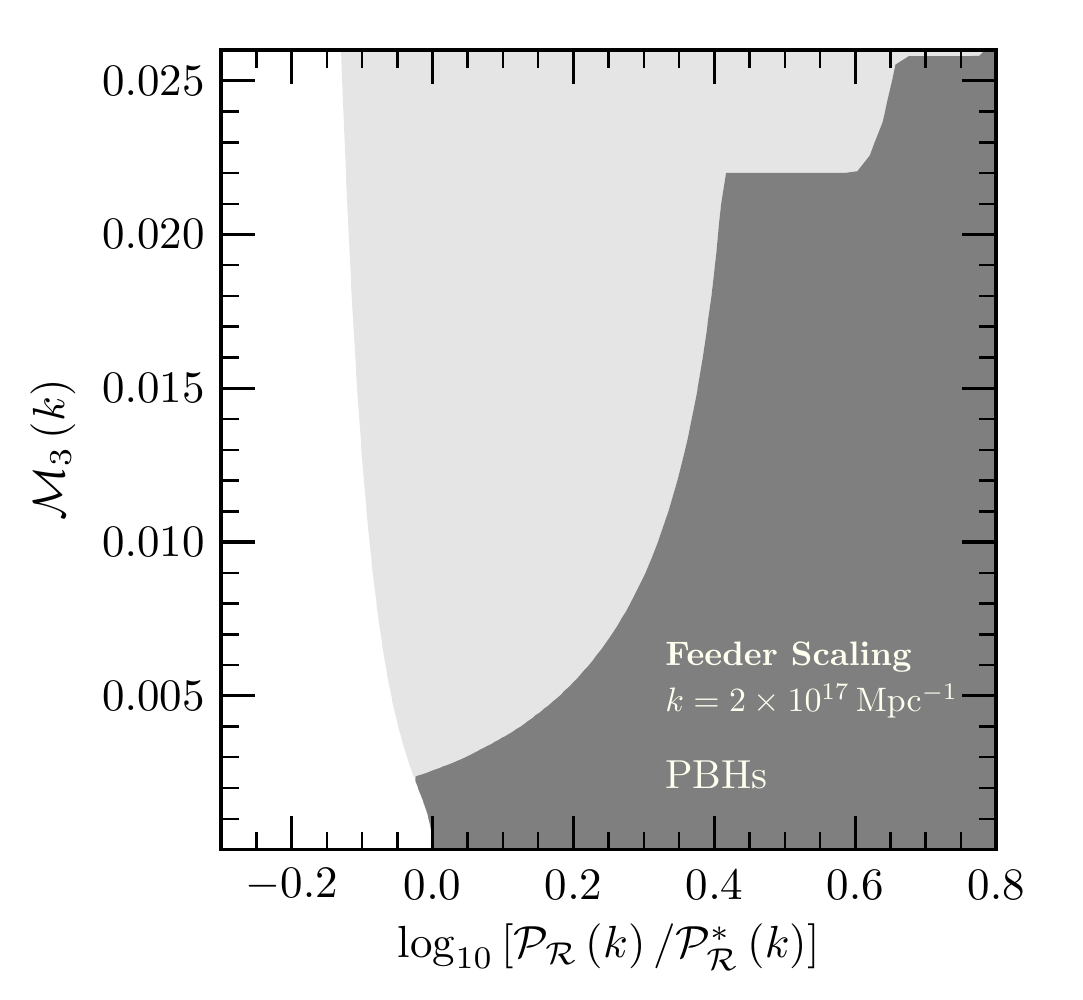} 
\caption{Bounds from UCMHs and PBHs on non-Gaussianity expressed as $\M_3(k)$ (the dimensionless skewness, defined in Eq.(\ref{eq:MnRs})), as a function of the amount of Gaussian power in curvature perturbations.  Dark shaded regions are ruled out at 95\% CL by observations.  Light shaded regions would have been ruled out by a naive application of the observational limits but correspond to areas where our analysis is invalid because our expansion in Eq.~(\ref{eq:finalbeta}) breaks down.  The limits are given at $k=2\times10^3$\,Mpc$^{-1}$ (top), $k=2\times10^7$\,Mpc$^{-1}$ (center) and $k=2\times10^{17}$\,Mpc$^{-1}$ (bottom), for both the feeder and hierarchical scalings.  Here the Gaussian power is expressed relative to the respective limits from PBHs and UCMHs on a Gaussian spectrum.  In general UCMHs probe non-Gaussianity at much lower overall amplitudes, but PBHs are sensitive to a broader range of scales.  At $k=2\times10^3, 2\times10^7, 2\times10^{17}$\,Mpc$^{-1}$, for PBHs $\log_{10}(\mathcal{P}_\mathcal{R}^*) = -2.65, -2.58, -2.91$.  For UCMHs with collapse redshift $\zc=1000$, $\log_{10}(\mathcal{P}_\mathcal{R}^*) = -6.54, -6.87$ at $k=2\times10^3, 2\times10^7$\,Mpc$^{-1}$, whereas with $\zc=200$, $\log_{10}(\mathcal{P}_\mathcal{R}^*) = -8.25, -8.47$.
\label{fig:UCMH+PBH}}
\end{center}
\end{figure*}

UCMHs were initially discussed in early papers by Berezinsky and collaborators \cite{Berezinsky:2003vn,Berezinsky:2005py,Berezinsky:2006qm,Berezinsky:2007qu}, and have been recently studied in the context of lensing signatures \cite{Ricotti:2009bs,Li12,Zackrisson:2012fa}, gamma-ray and neutrino signals from dark matter annihilation or decay \cite{Scott:2009tu,Lacki:2010zf,Josan:2010vn,Berezinsky:2010kq,Berezinsky:2010kr,Bringmann:2011ut,Yang:2011eg,Yang12b,Yang:2013dsa}, and impacts on the CMB \cite{Zhang:2010cj,Yang:2011jb,Yang:2011ef,Yang12b}.  To date the strongest limits come from gamma-ray observations \cite{Bringmann:2011ut}, indicating that $\ns<1.17$ for a scale-free spectrum, and $\mathcal{P}_\mathcal{R}(k)\lesssim 10^{-6}$--$10^{-7}$ for generalized spectra, on scales corresponding to $3\, \mathrm{Mpc}^{-1}<k<3\times10^{7}$\,Mpc$^{-1}$.  The corresponding direct limits on $\beta$ at the two example scales of Figure \ref{fig:UCMH+PBH} accessible with UCMHs are $\beta < 1.8\times10^{-9}$ ($k=2\times10^{3}$\,Mpc$^{-1}$) and $\beta < 3.0\times10^{-6}$ ($k=2\times10^{7}$\,Mpc$^{-1}$). 

To explore the impact of UCMHs on the allowed parameter space for non-Gaussianity, we have implemented Eq.\ (\ref{eq:finalbeta}) in the analysis framework of \cite{Bringmann:2011ut}, using combined 1-year \textit{Fermi}-LAT limits on the abundance of UCMHs from Galactic and extragalactic source searches\footnote{See Appendix \ref{appendix:ptsrc} for details of our slight improvement of the treatment of sources over \cite{Bringmann:2011ut}.}, as well as the overall diffuse gamma-ray flux in the direction of the Galactic poles.  Whilst 2-year source searches (e.g.\ \cite{Hans12}) and corresponding diffuse data are available, we conservatively adopt 1-year sensitivities to be certain that no dark matter sources were found in the observing period we consider.  The reader is referred to \cite{Bringmann:2011ut} for extensive discussion of this issue.  Our gamma-ray limits assume the canonical thermal cross-section for dark matter annihilation ($3 \times10^{-26}$\,cm$^3$\,s$^{-1}$) into $b$ quarks and anti-quarks, and a relatively conservative dark matter mass of 1\,TeV (smaller masses would result in stronger limits).

In Figure \ref{fig:UCMH+PBH}, we show the ability of UCMHs and PBHs to exclude different values of $\M_3$ for the two scaling scenarios, as a function of the departure of the underlying Gaussian power from the current Gaussian limit.  Dark shaded regions are excluded at 95\% confidence level (CL).  By expressing the underlying Gaussian power \emph{relative to present limits on Gaussian power} from each class of compact object, it is possible to compare the limits on $M_3$ from UCMHs and PBHs directly.  Although UCMHs provide a much stronger limit than PBHs on the Gaussian power (as known from \cite{Bringmann:2011ut}), PBHs provide a stronger lever arm for constraining non-Gaussianity below their Gaussian limit than UCMHs.  On the other hand, our expansion is well-behaved to larger $\M_3$ for UCMHS, allowing them to exclude stronger non-Gaussianities than PBHs.

The reference power $\mathcal{P}_\mathcal{R}^*$ that we have used for normalising the curves shown in Figure \ref{fig:UCMH+PBH} is the observational limit on the Gaussian power at each scale.  For PBHs at $k=2\times10^3, 2\times10^7, 2\times10^{17}$\,Mpc$^{-1}$, this corresponds to $\log_{10}(\mathcal{P}_\mathcal{R}^*) = -2.65, -2.58, -2.91$.  For UCMHs, the limit on the Gaussian power is much stronger.  Precisely how much stronger depends on the latest redshift at which UCMHs can be assumed to undergo gravitational collapse.  Assuming that only objects collapsing before a redshift of $\zc=1000$ form UCMHs, the reference power is $\log_{10}(\mathcal{P}_\mathcal{R}^*) = -6.54$ at $k=2\times10^3$\,Mpc$^{-1}$, and $\log_{10}(\mathcal{P}_\mathcal{R}^*) = -6.87$ at $2\times10^7$\,Mpc$^{-1}$.  

Setting $\zc=1000$ follows earlier work \cite{Ricotti:2009bs,Bringmann:2011ut}, and is a very conservative choice.  The defining characteristic of a UCMH is that it forms in isolation from a spherical collapse; the processes of radial collapse and secondary infall give UCMHs extremely steep dark matter density profiles, thereby distinguishing them from halos formed at late times.  Exactly when the approximation of radial infall breaks down is not yet known \cite{Bringmann:2011ut}, as this depends sensitively on the velocities of dark matter particles and interactions between growing overdensities.  More detailed simulation work is sorely needed to answer this question.  In principle UCMHs may form at lower redshifts, as long as the presence of different-sized overdensities does not spoil the assumption of spherical collapse.  The smaller the redshift, the more unlikely it is that this requirement will be fulfilled.  Later collapse redshifts allow smaller amplitude perturbations to form UCMHs, reducing $\del{UCMH}$ and increasing $\beta_\mathrm{UCMH}$ for a given combination of $\M_3$ and $\mathcal{P}_\mathcal{R}$.  If UCMHs continue to be formed as late as $\zc=200$, their non-observation begins to constrain non-Gaussianities on small scales, even if the power spectrum is near the level observed on CMB scales: $\log_{10}(\mathcal{P}_\mathcal{R}^*) = -8.25$ at $k=2\times10^3$\,Mpc$^{-1}$, and $\log_{10}(\mathcal{P}_\mathcal{R}^*) = -8.47$ at $2\times10^7$\,Mpc$^{-1}$.  We stress again however, that there exists \emph{no observational evidence whatsoever} that the power on small scales is actually this low (see Figure \ref{fig:PlanckPS}); even with a very conservative collapse redshift of $\zc=1000$ UCMHs therefore place significant new constraints on non-Gaussianity at small scales.

\section{Discussion \& Conclusions}

We have provided a general technique for using the abundance of any object to constrain weakly non-Gaussian statistics of the primordial fluctuations. We suggest precise criteria for the amplitude of non-Gaussianity that can be accurately constrained using Petrov-type expansions and emphasize that there is not a one-to-one mapping between the overall level of non-Gaussianity and the size of the skewness ($\mathcal{M}_3$). This is important when using the abundance of rare objects to constrain non-Gaussianity: number counts are sensitive to {\it any} deviation from Gaussianity, but a measured value of $\beta$ does not constrain a parameter like $f_{NL}$ (typically interpreted as the amount of skewness) without additional information about the higher moments of the PDF. 

In general PBHs provide stringent constraints on non-Gaussianities only in the presence of fluctuations with relatively large variance, from $\mathcal{P}_\mathcal{R}\gtrsim1-2\times10^{-3}$ at intermediate ($k=2\times10^3$\,Mpc${^{-1}}$) and small ($k=2\times10^7$\,Mpc${^{-1}}$) scales, to $\mathcal{P}_\mathcal{R}\gtrsim7\times10^{-4}$ at very small scales.  Notably though, at very small scales PBHs provide the only available limit on the allowed amount of non-Gaussianity. UCMHs provide stronger constraints, and so may be more likely to eventually also constrain non-Gaussianity, but do not reach as small scales as PBHs. 

For both UCMHs and PBHs, it is straightforward to map our constraints on $\mathcal{M}_3$ to constraints on a non-Gaussian parameter in the primordial fluctuations. For example, for the local ansatz $\mathcal{R}(x)=\mathcal{R}(x)_G+\frac{3}{5}f_{NL}[\mathcal{R}_G(x)^2-\langle\mathcal{R}_G(x)^2\rangle]$ the smoothed, integrated, dimensionless skewness $\mathcal{M}_3 \sim$ $\langle\mathcal{R}(x)^3\rangle_R/(\langle\mathcal{R}(x)^2\rangle_R)^{3/2}$ is very nearly equal to $f_{NL}\mathcal{P}_\mathcal{\mathcal{R}}^{1/2}$.  For this case, if the variance is large enough, the abundance of UCMHs constrains $f_{NL}\lesssim \mathcal{O}(100)$ and PBHs constrain $f_{NL}\lesssim \mathcal{O}(10)$ on smaller scales. However, it is important to remember that this does not mean UCMHs are less powerful -- quite the opposite. For the constraint from either object, the maximum value of $f_{NL}$ corresponds to $\mathcal{M}_3\sim\mathcal{O}(0.1)$, so those values of $f_{NL}$ correspond to equally non-Gaussian distributions. The difference in constraints phrased as a value of $f_{NL}$ comes from the difference in the variance of the fluctuations.  For reference, recall that on CMB scales, where the amplitude of fluctuations is currently measured, $f_{NL}\sim\mathcal{O}(10)$ corresponds to $\mathcal{M}_3\sim\mathcal{O}(10^{-3})$. 

\begin{figure}[tbp]
\includegraphics[height=65mm]{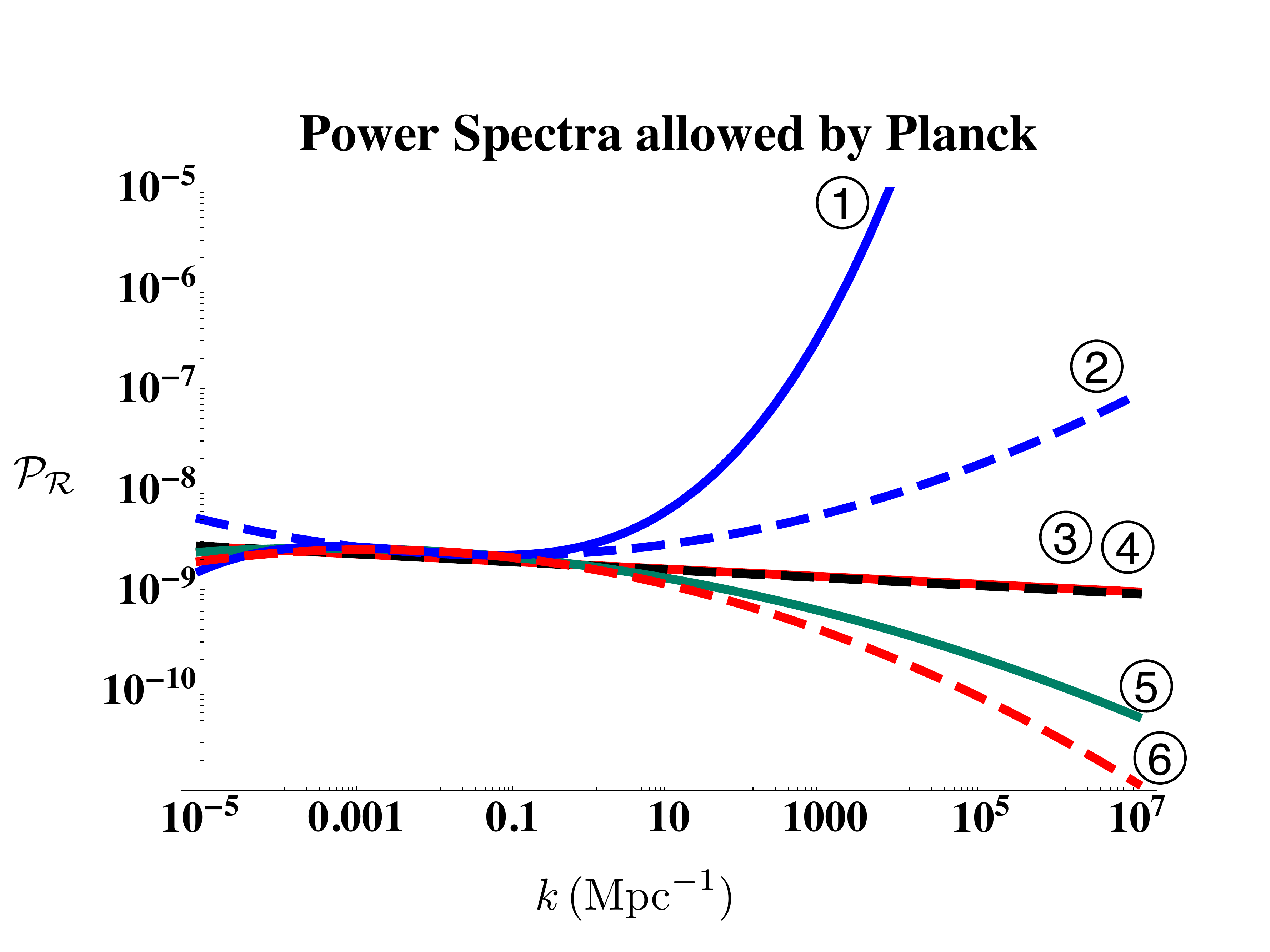} 
\caption{The power spectra corresponding to the best fit results of the {\it Planck} satellite for several choices of allowed parameters and combinations with other data sets \cite{PlanckInflation}.  The minimal 6 parameter ``$\Lambda$CDM" fit models the power spectrum amplitude ($A_s$) with a constant tilt ($n_s-1$). If additional scale dependence of $n_s$ and/or a non-zero amplitude of tensor modes ($r$) are allowed, the best fit power spectrum can be different. From top to bottom at large $k$, the allowed parameters and data sets shown (where WP = {\it WMAP} polarization data) are: (1) $\Lambda$CDM + $dn_s/d{\rm ln}k$ + $d^2(n_s)/d{\rm ln}k^2$, {\it Planck}+WP (solid blue); (2) $\Lambda$CDM + $r$ + $dn_s/d{\rm ln}k$, {\it Planck}+WP (dashed blue); (3) $\Lambda$CDM + $r$, {\it Planck}+WP (solid red); (4) $\Lambda$CDM, {\it Planck}+WP (dashed black); (5) $\Lambda$CDM + $dn_s/d{\rm ln}k$, {\it Planck}+WP (solid green); (6) $\Lambda$CDM + $r$ + $dn_s/d{\rm ln}k$, {\it Planck}+WP+BAO  (dashed red). {\it Planck} does not report the new best fit value of $A_s$ for the models with more than 6 parameters, but the 2$\sigma$ uncertainty in the amplitude for the minimal model is within the thickness of the lines. The $\Lambda$CDM and $\Lambda$CDM + $r$ models (lines 3 and 4) are reported at a pivot of $k_*=0.002$Mpc$^{-1}$, while the others have a pivot of $k_*=0.05$Mpc$^{-1}$.
\label{fig:PlanckPS}}
\end{figure}

To constrain non-Gaussianity, both PBHs and UCMHs require significantly more power than extrapolating the CMB spectrum from the standard six parameter fit would suggest, but PBHs require even more power than UCMHs. So, given a prediction for the power on small scales, UCMHs are more likely to constrain non-Gaussianity. In addition, current observational constraints on cosmological models that allow more than six parameters show that additional power on small scales is hardly ruled out. Figure \ref{fig:PlanckPS} shows the power spectra allowed by fits to the {\it Planck} satellite data \cite{PlanckInflation}. 

The constraints we quote here assume that the dark matter particle is a standard thermal relic with detectable annihilation products \cite{Bringmann:2011ut}. Unfortunately, near-future astrometric microlensing searches for UCMHs can only constrain $\beta_\mathrm{UCMH} \lsim 0.3$ \cite{Li12}.  For these large values of $\beta$, UCMHs are not sufficiently rare to be a useful probe of non-Gaussianity because $\nu_\mathrm{c} \simeq 1$, which is smaller than the crossing point between Gaussian and skewed PDFs.  Consequently, changing $\mathcal{M}_3$ does not significantly affect the abundance of relatively common objects $(\beta \gsim 0.1)$.

Aside from the abundances of rare objects, the only other probe of the small-scale inhomogeneities is the spectral distortion of the CMB, specifically the $\mu$-distortion, which is sensitive to the fluctuations on scales on the order of $k=50-10^4$ Mpc$^{-1}$\cite{Hu:1994bz,Chluba:2011hw,Chluba:2012we}. As with PBHs and UCMHs, CMB $\mu$-distortion can also constrain non-Gaussianity, but the nature of the constraints on both characteristics is qualitatively different. First, CMB $\mu$-distortion is sensitive to the total power in fluctuations over a wider range of scales, while the objects are sensitive to the power only very near the scale of the object, $k\sim1/R$. Second, $\mu$-distortion is sensitive to only the squeezed limit of the bispectrum rather than the total skewness. That is, this measurement can constrain the amplitude in the correlation of three momentum modes with $k_1\ll k_2\sim k_3$ \cite{Pajer:2012vz,Ganc:2012ae}. If non-Gaussianity follows the local ansatz (which has a strong signal in the squeezed bispectrum), a futuristic CMB probe like PIXIE \cite{Kogut:2011xw} could constrain $f_{NL}\sim\mathcal{O}(10^3)$ for a scale-invariant power spectrum and $f_{NL}\sim\mathcal{O}(10)$ for an enhanced small-scale power spectrum that saturates current bounds ($\mathcal{M}_3\sim0.04$ in both cases) \cite{Pajer:2012vz}. As can be seen from the top two panels in Figure \ref{fig:UCMH+PBH}, the $\mu$-distortion bound would be about an order of magnitude better than the bounds from object counts at that scale if the non-Gaussianity is local and hierarchical. The bounds may be similar for non-Gaussianity with a stronger, feeder type scaling and a local bispectrum. The $\mu$-distortion probe is then quite complementary to the abundance of PBHs and UCMHs in terms of the range of scales probed and the relationship between the observables and the primordial fluctuations.

\acknowledgments
J. Yana Galarza and S. Shandera were supported in the early stages of this work by the Perimeter Institute for Theoretical Physics. Research at Perimeter Institute is supported by the Government of Canada through Industry Canada and by the Province of Ontario through the Ministry of Research and Innovation. S. Shandera has also been supported by the Eberly Research Funds of The Pennsylvania State University. The Institute for Gravitation and the Cosmos is supported by the Eberly College of Science and the Office of the Senior Vice President for Research at the Pennsylvania State University. A. Erickcek supported in part by the Canadian Institute for Advanced Research. P. Scott is supported by the Banting program, administered by the Natural Science and Engineering Research Council of Canada.

\appendix
\section{Generalized source limits on UCMHs}
\label{appendix:ptsrc}
Our analysis here actually includes a slightly improved form of the point source analysis of \cite{Bringmann:2011ut}.  Following from Eq.\ (18) in \cite{Bringmann:2011ut}, if there is a single UCMH present somewhere in the Milky Way, we identify the probability $P^\mathrm{\,seen}_{1}$ of actually seeing that UCMH as
\begin{equation} 
\label{P_1_better}
P^\mathrm{\,seen}_{1} = xP^\mathrm{\,exists}_{d<d_\mathrm{obs},1} = x\frac{M_{d<d_\mathrm{obs}}}{M_\mathrm{MW}}.
\end{equation}
Here $d$ is the true distance of the UCMH from Earth, $d_\mathrm{obs}$ is the maximum observable distance required for detection at CL $x$, $P^\mathrm{\,exists}_{d<d_\mathrm{obs},1}$ is the probability of the UCMH residing within distance $d_\mathrm{obs}$ of Earth, $M_\mathrm{MW}$ is the dark mass of the Milky Way and $M_{d<d_\mathrm{obs}}$ is the dark mass within $d_\mathrm{obs}$. 

This leads to an improved version of Eq.\ (24) in \cite{Bringmann:2011ut}, which gives the maximum allowed fraction of matter in UCMHs today
\begin{equation}
\label{fmax_better}
f_\mathrm{max} = \frac{f_\chi M^0_\mathrm{UCMH}}{M_\mathrm{MW}}\frac{\log(1-y)}{\log(1-x\frac{M_{d<d_\mathrm{obs}}}{M_\mathrm{MW}})}.
\end{equation}
Here $f_\chi \equiv \rho_\chi/\rho_\mathrm{m}$ is the fraction of matter in the Universe that is dark, $M^0_\mathrm{UCMH}$ is the present-day UCMH mass, and $y$ is the desired CL of the limit $f_\mathrm{max}$.  The relationship between $\beta$ and $f_\mathrm{max}$ can be found in Eq.\ (21) of \cite{Bringmann:2011ut}. Eq.\ (\ref{fmax_better}) is valid for all $x,y<1$.  For $1-y \gg 1-x$, it agrees with Eq.\ (24) in \cite{Bringmann:2011ut}; whilst both forms are valid for the values of $x$ and $y$ we use here ($1-y=5\times10^{-2}$, $1-x=6\times10^{-7}$, as in \cite{Bringmann:2011ut}), the latter breaks down as $y\to x$, so Eq.\ (\ref{fmax_better}) is more general.

\providecommand{\href}[2]{#2}\begingroup\raggedright\endgroup

\end{document}